\documentclass[11pt,a4paper]{article}
\usepackage{jheppub}

\usepackage{amssymb,amsmath,amstext,amsfonts,empheq}
\usepackage{bbold}
\usepackage{bm}
\usepackage{graphicx}
\usepackage{xcolor}
\usepackage{overpic}
\usepackage{subcaption}
\usepackage{braket}
\usepackage{tcolorbox}

\newcommand{\be}{\begin{equation}}
\newcommand{\ee}{\end{equation}}
\newcommand{\bea}{\begin{eqnarray}}
\newcommand{\eea}{\end{eqnarray}}

\newcommand{\bal}{\begin{aligned}}
\newcommand{\eal}{\end{aligned}}

\newcommand{\Mp}{M_{\rm Pl}}

\newcommand{\etaperp}{\eta_\perp}

\newcommand{\R}{\mathcal{R}}
\newcommand{\F}{\mathcal{F}}
\newcommand{\Rh}{\hat{\mathcal{R}}}
\newcommand{\Fh}{\hat{\mathcal{F}}}

\newcommand{\kf}{k_{\textrm{out}}}

\newcommand{\bog}{Bogoliubov }

\newcommand{\U}{\mathcal{U}}
\definecolor{lightgray}{rgb}{0.83, 0.83, 0.83}

\newcommand{\bk}{\boldsymbol{k}}
\newcommand{\bp}{\boldsymbol{p}}
\newcommand{\bq}{\boldsymbol{q}}
\definecolor{alizarin}{rgb}{0.82, 0.1, 0.26}

\newcommand{\Omegagw}{\Omega_{\mathrm{GW}}}

\newcommand{\di}{{\rm d}}

\newcommand{\p}{\mathcal{P}}

\usepackage{soul}
\definecolor{lgray}{gray}{0.90}

\newcommand{\bx}{\boldsymbol{x}}
\definecolor{alizarin}{rgb}{0.82, 0.1, 0.26}

\newcommand{\tauf}{\tau_{\rm out}}
\newcommand{\zf}{z_{\mathrm{out}}}

\definecolor{forestgreen(web)}{rgb}{0.13, 0.55, 0.13}

\newcommand{\cG}{\mathcal{G}}

\newcommand{\bs}{\left\vert \beta_* \right\vert}

\def\ga{~\mbox{\raisebox{-.6ex}{$\stackrel{>}{\sim}$}}~}

\title{ \centering
 \huge
One-loop infrared rescattering by enhanced scalar fluctuations during inflation
}
\author[a]{Jacopo Fumagalli,}
\author[b,c]{Sukannya Bhattacharya,}
\author[b,c]{Marco Peloso,}
\author[d]{S\'ebastien Renaux-Petel}
\author[d]{Lukas T. Witkowski}

\affiliation[a]{Departament de F\'isica Quàntica i Astrofísica i 
Institut de Ciències del Cosmos, \\[1mm]
Universitat de Barcelona,\\[1mm] Martí i Franquès 1, 08028 Barcelona, Spain}

\affiliation[b]{Dipartimento di Fisica e Astronomia “Galileo Galilei”, Universit\`a di Padova, 35131 Padova, Italy}

\affiliation[c]{INFN, Sezione di Padova, 35131 Padova, Italy}

\affiliation[d]{Institut d’Astrophysique de Paris, GReCO, UMR 7095 du CNRS et de Sorbonne Universit\'e,
98 bis bd Arago, 75014 Paris, France}

\emailAdd{jfumagalli@fqa.ub.edu}
\emailAdd{sukannya.bhattacharya@unipd.it}
\emailAdd{marco.peloso@pd.infn.it}
\emailAdd{renaux@iap.fr}
\emailAdd{lukas.t.witkowski@gmail.com}

\abstract{We show that, whenever the perturbations of some field are excited during inflation by a physical process on sub-horizon scales, they unavoidably generate, even through gravitational interactions alone, a significant resonant IR cascade of power down to scales that are of the order of the horizon at that time (we denote these scales as near IR). We provide general analytic one-loop results for the enhancement of the IR power of the curvature perturbation generated by this effect, highlighting the role played by the resonance. We then study a number of examples in which the excited state is:  (i) an isocurvature field, (ii) the  curvature perturbation itself, (iii) a mixture of curvature and isocurvature fluctuations driven to an excited state by their coupled dynamics. In the cases shown, the cascade significantly modifies the near IR part of the power spectrum of the curvature perturbation with respect to the linear theory, indicating that this effect can impact the phenomenology associated with a variety of mechanisms considered in the literature, notably concerning primordial black holes and gravitational waves. 
}
\begin{document}
\hfill{\flushright {}}
\maketitle

\section{Introduction}

Recent and future advances in astrophysical and cosmological observations are opening new observational windows at small scales that may allow us to test the inflationary history at later times than those presently probed by Cosmic Microwave Background (CMB) and Large Scale Structure (LSS) observations. A number of physical processes in act during these times can lead to enhanced scalar and/or tensor fluctuations, that in turn could lead to the production of primordial black holes (PBH) \cite{Ivanov:1994pa,Garcia-Bellido:1996mdl}, with a number of observational prospects \cite{Sasaki:2018dmp,Carr:2020gox,Carr:2020xqk,Green:2020jor,Carr:2023tpt}, and/or the generation of a detectable stochastic background of gravitational waves (SGWB) \cite{Caprini:2018mtu,Guzzetti:2016mkm,LISACosmologyWorkingGroup:2022jok}. 

In light of the several present (LIGO, Virgo, KAGRA, IPTA) or upcoming (for instance LISA, Einstein Telescope, Cosmic Explorer, SKA) opportunities to detect these probes, recent literature has proliferated with mechanisms capable of enhancing primordial scalar fluctuations on short scales. In this context, there has been very recently an intense discussion on loop corrections induced on the very far infrared (IR) regime \cite{Kristiano:2022maq,Riotto:2023hoz,Kristiano:2023scm,Riotto:2023gpm,Firouzjahi:2023aum,Firouzjahi:2023ahg,Franciolini:2023lgy,Tasinato:2023ukp,Cheng:2023ikq} (such as the CMB scales) by enhanced scalar perturbations of much higher comoving momentum $k$.
Hints for significant one-loop contribution in the far IR scales $k \equiv k_{\mathrm{IR}, \,\mathrm{far}}$, to the point of undermining the consistency of all these set-ups, have appeared in computations including only a subset of vertices as derived from the interaction Hamiltonian. However, it has been recently shown that these effects disappear (due to cancellations) once all relevant interaction terms are taken into account \cite{Fumagalli:2023hpa}.

In the present work, we point out that, despite this, the enhancement of perturbations at some scale $k_*$ indeed often leads to significant and observable IR production, however, at momentum scales $k_{\mathrm{IR},\,\mathrm{near}}$ that are only a few orders of magnitude smaller than $k_*$, down to the horizon scale. To our knowledge, the importance of this kind of effects was first pointed out in the work \cite{Barnaby:2009mc}, and further explored in some of the following literature~\cite{Barnaby:2009dd,Barnaby:2010ke,Barnaby:2010sq,Pearce:2017bdc}, where the \textit{rescattering} of the excited modes among themselves (namely, the nonlinear interactions) and against the inflaton zero mode result in an IR cascade that can transfer power from shorter to larger scales. While Ref. \cite{Barnaby:2009mc} considered the specific case of an enhancement due to the $\left( \phi - \phi_0 \right)^2 \chi^2$ interaction between the inflaton $\phi$ and another field $\chi$, which is suddenly produced when the inflaton reaches the value $\phi_0$ during inflation, we point out here that this IR production is a much more general effect, that can take place just by gravitational interactions (even in the absence of any other direct coupling) whenever a field is suddenly excited at sub-horizon scales at some given time $\tauf$ during inflation. This can be the result of a variety of physical processes, such as a momentary change in the speed of the inflaton \cite{Starobinsky:1992ts,Kaloper:2003nv,Ashoorioon:2006wc,Bean:2008na,Ashoorioon:2017toq,Ashoorioon:2018uey,Ballesteros:2018wlw,Tasinato:2020vdk,Dalianis:2021iig,Inomata:2021tpx} or a sharp turn in multifield space \cite{Achucarro:2010da,Palma:2020ejf,Fumagalli:2020adf,Fumagalli:2020nvq,Braglia:2020taf,Iacconi:2021ltm,Bhattacharya:2022fze,Aragam:2023adu} or also multiple-stage inflation~\cite{Polarski:1992dq,Adams:1997de,Pi:2017gih,Pi:2019ihn,DAmico:2020euu,DAmico:2021vka}. Depending on the specific model, the excited field could be an entropic (i.e. isocurvature) degree of freedom, or the adiabatic mode itself. In the latter case we label the effect as \textit{self-scattering}.   

While the precise amount of IR production from these excited states is model-dependent, we show that this effect generally occurs under the above conditions. To be specific, we consider the situation in which $k_*$ is a few orders of magnitude greater than the comoving Hubble rate at the time $\tauf$. For simplicity, we assume that the excitation is sufficiently large so that these modes can be treated as a classical source.\footnote{In technical terms, this corresponds to large occupation numbers, in a regime in which the quantum commutators between these sourcing fields can be neglected, as we compute below.} We then show that the model-independent and ever-present gravitational interactions among these modes are already enough to induce a large production in the near IR region, which is peaked at scales that are of the size of the horizon at $\tauf$. 

We show that this effect is well captured by analytic perturbation theory. We denote the excited sourcing modes of comoving momenta $k_*$ as the linear field(s), as in the specific concrete examples that we consider, the sudden production of these modes is described by linear physics (since this generation occurs at sub-horizon scales, we can think of it as a process of particle production). To leading order in the gravitational interaction, IR rescattering processes can then be thought as $2 \to 1$ processes where two incoming quanta of momenta $k_*$ produce a mode of momentum $k_{\mathrm{IR,\,near}}$. The corresponding correction to the power spectrum of the sourced mode can then be described by a one-loop diagram. These contributions are under control at the scale $k_*$ (ensuring the consistency of the perturbative approach) and are negligible at $k_{\mathrm{IR,\,far}}$ scales, but they can instead be significant at the $k_{\mathrm{IR,\,near}}$ scales.

Our analytical results clarify why rescattering from modes inside the horizon is so effective. The sourcing modes with sub-horizon momenta $k_*$ behave as classically oscillating waves, giving rise to a resonant amplification of the sourced IR modes. An analogous resonant effect was recently shown to occur in the production of GW from excited sub-horizon scalar modes~\cite{Fumagalli:2021mpc}. This bears strong similarities with the production of scalar perturbations studied in this work. One main difference is that in the case of scalar perturbations, additional nonlinear production might occur from the dynamics (and, possibly, the interactions) responsible for the excitation of the sourcing field, or from any coupling that might exist between these fields besides the gravitational one. As already mentioned, the goal of the present work is to emphasize the universality of these effects, and therefore we only compute the rescattering due to gravitational interactions after the excitation of the sourcing fields has taken place. Since we do not expect cancellations between different classical nonlinear processes, our results should be understood as a lower estimate of the IR power produced in the scalar sector. Signatures in specific models might be greater than our estimates, and they should be computed on a case-by-case basis. We see that already for the gravitational interactions considered in this work the near IR one-loop contribution to the primordial power spectrum can ``emerge" aside to the ``primary" tree-level ultraviolet (UV) peak, as we show explicitly, for instance, in Figure \ref{PS-etajump} and, in an even more striking manner, in Figure \ref{multi_benchmark}. This additional IR contribution can have important observable consequences for PBH and GW observations.

The manuscript is organized as follows. In Section \ref{spectator}, we lay out the formulation to describe the linear sourcing fields, and provide some of their properties that are then employed in the following sections. 

In Section~\ref{sec:spectator2} we compute the one-loop contribution to the power spectrum of the curvature perturbation 
from an excited isocurvature field. Specifically, in Subsection~\ref{sec:FFR} we formulate the third-order action describing the gravitational interactions between two entropic and one adiabatic mode in a form adapted to our loop computation. In Subsection~\ref{subsec:inin} we employ the in-in formalism to derive the formal expression that provides the one-loop correction to the adiabatic mode. In Subsection~\ref{subsec:classical} we demonstrate the equivalence between the in-in formalism and the classical Green function method in the case of large occupation numbers (showing explicitly that the source can be taken as a classically commuting field in this limit, see also Ref.~\cite{Barnaby:2011qe} for similar considerations in a different mechanism, and \cite{Inomata:2022yte} for a recent use of the full equivalence for polynomial interactions). In Subsection \ref{subsec:1loop} we evaluate the one-loop power spectrum and show how our analytical solution captures the resonant effect responsible for the large IR production. In Subsections \ref{subsec:analytic-peaked-specfield} and \ref{subsec:UVsource} we show how this result simplifies in, respectively, the case of narrowly peaked and deep sub-horizon sourcing field. This section is significantly more detailed than the two following ones, as a large number of concepts and results developed here continue to apply, with only minor modifications, in the following cases. 

In Section \ref{sec:single} we consider the situation where the curvature perturbation itself is in an excited state and compute the rescattering due to its self-interactions. While the general formal expression for the one-loop production is worked out in Subsection \ref{subsec:general-self}, the following Subsection \ref{subsec:examplesingle} studies an explicit example in which the excitation of the adiabatic mode is due to a phase of large slow-roll parameter $\eta$. 

Section \ref{sec:multifield} studies the more generic multifield set-up where both entropic and adiabatic fluctuations are excited at $\tauf$. This is also divided in a first Subsection \ref{subsec:general-multi} where we work out the general formal expression, and in a second Subsection~\ref{subsec:examplemulti} where we study the specific example of the excitation due to a strong turn in field space. 

We conclude and mention possible outlooks in Section \ref{sec:conclusion}.

\section{Linear fields}\label{spectator}

We consider the Lagrangian for the curvature perturbation and one massless isocurvature field $\F$ in standard multifield scenarios, gauging to zero the adiabatic field, so that the spatial part of the metric reads $g_{ij}=a^2 e^{2 \R} \delta_{ij}$. The second order action for $\R$ and $\F$ is given by
\begin{align}\label{quadraticR}
\mathcal{L}_{\R}^{(2)}&=\frac{1}{2} (\Mp^2 a^2 2\epsilon) \left[ (\R')^2 - (\partial_i\R)^2 \right],
\end{align}
and 
\be\label{quadraticF}
\mathcal{L}_{\F}^{(2)}=\frac{1}{2} a^ 2  \left[ (\F')^2 - (\partial_i\F)^2 \right],
\ee 
where prime means derivative with respect to the conformal time $\tau$, $\epsilon \equiv -\dot{H}/{H^2}$ is the usual slow-roll parameter (with dot denoting derivative with respect to physical time), $H \equiv \frac{\dot{a}}{a}$ the Hubble rate, and $a$ is the scale factor. Note that the variable $\R$ is dimensionless, while $\F$ has mass dimension one. 

In momentum space, canonical quantization of the free fields evolving with the quadratic action leads to\footnote{For simplicity, we do not consider, in this section, the quantum mixing that can arise due to the interaction between $\R$ and other degrees of freedom before the transition. See sec. \ref{sec:multifield} for the general case.}

\begin{align}\label{interactionfield}
\Rh(\bp,\tau) =\R(p,\tau) \hat{a}_1(\bp) + \R^*(p,\tau) \hat{a}_1^{\dagger}(-\bp),\qquad
\Fh(\bp,\tau) =\F(p,\tau) \hat{a}_2(\bp) + \F^*(p,\tau) \hat{a}_2^{\dagger}(-\bp) \;, 
\end{align}
with 
\be\label{quantiz}
[\hat{a}_i(\bp),\hat{a}_j^{\dagger}(\bp')]= (2\pi)^3 \delta_{ij}\delta(\bp -\bp').
\ee
The mode functions $\R$ and $\F$ are solutions of the linear equations of motion (eom):
\be\label{eom}
-\frac{1}{2 \Mp^ 2}\frac{\delta \mathcal{L}^{(2)}}{\delta \R} \equiv (a^ 2 \epsilon \mathcal{\R}')' - a^ 2 \epsilon  \, \partial^2\R = 0, \qquad -\frac{\delta \mathcal{L}^{(2)}}{\delta \F} 
\equiv (a^ 2 \mathcal{\F}')' - a^ 2  \partial^2\F=0.
\ee
In a slowly changing inflationary background, and to leading order in slow-roll, the mode functions of the two fields are given, respectively, by
\be\label{excited1}
\R(p,\tau) = \frac{H}{\sqrt{2 p^ 3 \,2\epsilon \Mp^2}}\bigg[ \alpha_{\R}(p) \U(p\tau)+ \beta_{\R}(p) \U^* (p\tau)\bigg] \equiv \frac{H}{\sqrt{2 p^ 3 2\epsilon \Mp^2}} \widetilde{\R},
\ee
and
\be\label{excited2}
\F(p,\tau) = \frac{H}{\sqrt{2 p^3 }}\bigg[ \alpha_{\F}(p) \U(p\tau)+ \beta_{\F}(p) \U^* (p\tau)\bigg] \equiv \frac{H}{\sqrt{2 p^3}} \widetilde{\F},
\ee
where we introduced
\be\label{buildingmode}
\U(p \tau)=(1 + i p \tau) e^{-i p \tau}. 
\ee 
The quantization enforces that modes satisfy the  Wronskian normalization 
\be
W_\tau[\R,\R^*] \equiv \R (k,\tau) (\R^* (k,\tau))' - \R' (k,\tau)\R^* (k,\tau) = \frac{i}{2 \Mp^2 \epsilon \, a (\tau)^2}, \quad W_\tau[\F,\F^*] =\frac{i}{a  (\tau)^2}.
\ee
Plugging eqs. \eqref{excited1} and  \eqref{excited2} in these conditions leads, for both fields, to  
\be
|\alpha_i|^2-|\beta_i|^2 = 1 \;\;,\;\; i=\{\R,\F\}.
\label{quantization}
\ee
Let us now evaluate the commutators of these linear fields, as we need them in the following subsection  
\begin{align} 
\left[ \Rh \left(  \bk , \, \tau \right) ,\,  \Rh \left(  \bk',\, \tau'  \right)  \right] =&
\left( 2 \pi \right)^3 \delta^{(3)} \left( \bk + \bk' \right) 
\frac{H^2}{4 k^3 \epsilon \Mp^2} 
\left[ \U \left( k \tau \right) \U^* \left( k \tau' \right) - 
\U^* \left( k \tau \right) \U \left( k \tau' \right) \right] \,, \nonumber\\ 
\left[ \Fh \left(  \bk,\, \tau  \right) ,\,  \Fh \left(  \bk',\, \tau'  \right)  \right] =&
\left( 2 \pi \right)^3 \delta^{(3)} \left( \bk + \bk' \right) 
\frac{H^2}{2 k^3  } 
\left[ \U \left( k \tau \right) \U^* \left( k \tau' \right) - 
\U^* \left( k \tau \right) \U \left( k \tau' \right) \right] \,. \nonumber\\ 
\end{align}  
We can rewrite them in terms of the Green function associated with the operator appearing in the linear eom \eqref{eom}
\begin{align}
g_k \left( \tau ,\, \tau' \right) =&  \frac{\left( 1 + k^2 \tau \tau' \right) \sin \left[ k \left( \tau - \tau' \right) \right] - k \left( \tau - \tau' \right) \cos \left[ k \left( \tau - \tau' \right) \right] }{k^3 \tau^{'2}} \nonumber\\
=& i \, \frac{\U \left( k \tau \right) \U^* \left( k \tau' \right) - \U^* \left( k \tau \right) \U \left( k \tau' \right) }{2 k^3 \tau^{'2}} \,. 
\label{RR-FF}
\end{align} 
Using this expression, and noting that the Green function satisfies $\tau^{\prime 2} g_k \left( \tau ,\, \tau' \right) = - \tau^2 g_k \left( \tau' ,\, \tau \right)$, we rewrite the commutators as 
\begin{align}
\left[ \Rh \left( \bk,\,  \tau  \right) ,\,  \Rh \left( \bk', \, \tau'  \right)  \right] =&
\left( 2 \pi \right)^3 \delta^{(3)} \left( \bk + \bk' \right) \frac{i}{2 \epsilon \Mp^2 a^2 \left( \tau \right)} \, g_k \left( \tau' ,\, \tau \right)  \nonumber\\ 
\left[ \Fh \left( \bk, \, \tau  \right) ,\,  \Fh \left(  \bk' ,\, \tau' \right)  \right] =&
\left( 2 \pi \right)^3 \delta^{(3)} \left( \bk + \bk' \right)  \frac{i}{ a^2 \left( \tau \right)} \, g_k \left( \tau' ,\, \tau \right)  \,.
\label{commutator}
\end{align}
We stress that the commutators are independent of the value of the \bog coefficients, namely they acquire the same values for the vacuum and for the excited states. 

As mentioned in the introduction, we are interested in computing the one-loop contribution to the primordial power spectrum from a dynamically generated excited state during inflation. At early times and in the deep UV regime, the modes are initialized in the so called Bunch-Davies vacuum, i.e. $\alpha_{i} = 1$ and $\beta_{i}=0$ for $i=\{\R,\F\}$, so that the vacuum associated with the annihilation/creation operator in the expansion \eqref{interactionfield} is the one minimizing the Hamiltonian in the infinite past. We then assume that a sharp transition or an interaction takes place around a time $\tauf$ during inflation, resulting in an excited state. For simplicity, we start in Section \ref{sec:spectator2} by considering the case where only the entropic field is excited. We assume that, for $\tau>\tauf$ the linear dynamics goes back to a standard quasi de Sitter evolution, so that the entropic field mode functions are described by the expression \eqref{excited2},\footnote{In eq. \eqref{excited2} we neglect the entropic field fluctuation mass, since, as we show later,
the modes $\F$ that are relevant for our effect are in the sub-horizon regime, where a mass term $\leq H$ can be neglected. A small entropic mass $m_s\simeq O(H)$ is understood, to ensure that this field gives a negligible direct contribution to the energy density and perturbations at the end of inflation.} but now with $\beta_{\F} \neq 0$, and actually $|\beta_{\F}|\gg 1$ for a given range of scales. 
After the transition, $|\beta_{\F}(k)|^2$ is interpreted as the number density of quanta of the entropic field of comoving momentum of magnitude $k$. Our purpose is to obtain (via a one-loop computation) the unavoidable imprint that an excited $\F$ state induces on the field $\R$. Clearly, the full effect on $\R$ is model-dependent, as it changes according to the specific interaction between the two fields (as well as the one between $\R$ and any other degree of freedom that might have been relevant at $\tauf$). We thus focus on the only guaranteed model-independent contribution, namely the gravitational interaction between the two fields after the excitation of $\F$ has taken place. We disregard any additional effect; for instance we first assume, for simplicity, that the transition or the interactions that have excited $\F$ leave the field $\R$ in its vacuum state (up to negligible contributions). We therefore set $\alpha_{\R} = 1,\, \beta_{\R} = 0$, and, for brevity, we omit the suffix $\F$ from the \bog coefficients of the entropic field. Moreover, we disregard any possible direct (i.e non-gravitational) coupling between the two fields after the transition. This leaves us with the minimal unavoidable effect that the transition / interaction that took place at $\tauf$ has on $\R$.

\section{One-loop power spectrum boosted by an excited isocurvature field}\label{sec:spectator2}

We want to compute at one-loop the contribution to the power spectrum of $\R$ induced by its gravitational interaction with the excited isocurvature field $\F$. Equal-time correlators are computed with the in-in formalism leading to the master formula \cite{Schwinger:1960qe,Jordan:1986ug,Calzetta:1986ey}
\be
\langle \R^{n}(t)\rangle=\langle0| \left[\bar{T}\left(e^{i\int_{-\infty_+}^{\tau}d\tau'\mathcal{H}_{I}(\tau')}\right)\right]\R_{I}^{n}(\tau)\left[T\left(e^{-i\int_{-\infty_-}^{\tau}d\tau''\mathcal{\mathcal{H}}_{I}(\tau'')}\right)\right]|0\rangle ,
\ee
where the subscript $I$ labels fields in the interaction picture, namely fields evolving with the quadratic Hamiltonian and $\mathcal{H}_I$ is the interaction Hamiltonian.
Since the interaction we are interested in becomes relevant only after a preferred time $\tauf$, the lower extrema of integration can be trivialized and there are no subtlety regarding the $-\infty_\pm = -\infty(1\pm i \epsilon)$ prescription. In this case, one can rewrite the previous expression in terms of a series of nested commutators as follows: 
\begin{equation}
\langle\R^{n}(\tau)\rangle=\sum_{j=0}^{\infty}i^{j}\int_{\tauf}^{\tau}d\tau_{1}\int_{\tauf}^{\tau_{1}}d\tau_{2}..\int_{\tauf}^{\tau_{j-1}}d\tau_{j}\langle[\mathcal{H}_{I}(\tau_{j}),[\mathcal{H}_{I}(\tau_{j-1}),..[\mathcal{H}_{I}(\tau_{1}),\R_{I}^{n}(\tau)]..]\rangle.
\end{equation}
Our final goal is to compute the one-loop correction to the primordial power spectrum $\mathcal{P}_\R$ defined as
\be\label{defpower}
\langle \hat{\R}(\bk,\tau) \hat{\R}\left( \bk',\tau \right)\rangle = \left( 2\pi \right)^3 \delta \left( \bk + \bk' \right)  \frac{2\pi^2}{k^3}\mathcal{P}_\R(k,\tau).
\ee
We are thus interested in one-loop diagrams with two external $\Rh$-fields obtained via the insertion of two cubic interactions.~\footnote{As the nonlinear mode $\Rh$ produced by the rescattering is incoherent with the linear one, the one loop contribution to the power spectrum adds up with the linear one, with no interference term.}  
For that we need the cubic interaction Hamiltonian $\mathcal{H}^{(3)}= -\int d\bx \,\mathcal{L}^{(3)}$ where $\mathcal{L}^{(3)}$ is the cubic Lagrangian in the field perturbations.
In particular, the rescattering studied in this work is due to excited field(s) running in the loop, namely here the isocurvature field $\F$.
For this reason, we are interested in $\mathcal{L}^{(3)}\propto \F\F\R$.~\footnote{Cubic (respectively, quartic) gravitational interactions are proportional to $\frac{1}{\Mp}$ (respectively, $\frac{1}{\Mp^2}$).  As a consequence, a diagram with a single quartic vertex arising from $\mathcal{L}^{(4)}\propto \F\F\R\R$ scales with $\Mp$ in the same way as the diagrams with two cubic vertices that we are considering here. In general, all these contributions should be included. In the regime of high excitation of the spectator field that we are considering, $\left\vert \beta \right\vert^2 \gg 1$, we can, however, neglect the contribution of this diagram, since it is proportional to the expectation value of only two powers of $\F$, and therefore its amplitude is proportional to two fewer powers of $\beta$.}

\subsection{Third-order action}
\label{sec:FFR}

In general, the third order action can be written in the following form 
\be
\label{fullthird}
S^{(3)} =\int d^ 4 x \left( \mathcal{L}^{(3)}_{\mathrm{bulk}} + \mathcal{L}^{(3)}_{\partial} + \mathcal{L}^{(3)}_{\mathrm{eom}}\right) \;, 
\ee
where $\mathcal{L}^{(3)}_{\partial}$ and $\mathcal{L}^{(3)}_{\mathrm{eom}}$ label boundary terms and terms proportional to the linear equations of motion respectively, and the left-over $\mathcal{L}^{(3)}_{\mathrm{bulk}}$ is referred to as the ``bulk Lagrangian". Note that, in the boundary part of the action, it is sufficient to keep only total temporal derivative terms, as total spatial derivatives never play a role.

Importantly, we note that the split between the three contributions in \eqref{fullthird} is far from unique, with different forms of the action better suited for different purposes. The generic cubic action involving adiabatic and entropic fluctuations was derived in \cite{Garcia-Saenz:2019njm}, where careful manipulations have been made such that the forms of the three contributions in \eqref{fullthird} make manifest the genuine size of cubic interactions, as relevant for the computation of the primordial bispectrum, generalizing Maldacena's computation in the single-field case \cite{Maldacena:2002vr}. 

As explained above, here we only consider the model-independent gravitational interaction between the entropic/isocurvature and adiabatic/curvature fluctuations. This corresponds to the bulk term
\begin{align}\label{third_spectator}
\mathcal {L}^{(3)}_{\mathrm{bulk}} = & \frac{1}{2}\epsilon a^2\left( (\F')^2+ (\partial_{i}\F)^2 \right)\R -\epsilon a^2\F'\partial_i\F\partial_i\partial^{-2}\R'\,,
\end{align}
the boundary term 
\be
\label{forgotten}
\mathcal{L}_\partial = \frac{d}{d\tau}\left(-\frac{a}{2H} \R (\F')^ 2 \,\, + \,\, \textrm{fields\, only} \right) \;, 
\ee
and the equation of motion contribution
\begin{align}\label{eomterm}
\mathcal{L}_{\mathrm{eom}}^{(3)} =
- \frac{1}{a H} \frac{\delta \mathcal{L}^{(2)}}{\delta \F} \F' \R \,.
\end{align}
Here, $\delta \mathcal{L}^{(2)}/\delta  \F $ is defined in \eqref{eom}, and the contributions not explicitly written in \eqref{forgotten} only contain the fields $\F$ and $\R$, but not their time derivatives.

From this, a few comments are in order. Firstly, we recall that terms proportional to the linear equations of motion do not contribute to correlation functions. This is true in general for any diagram at any order in perturbation theory simply because the operators $\delta \mathcal{L}^{(2)}/\delta  \R ,\, \delta \mathcal{L}^{(2)}/\delta  \F$ are exactly zero when evaluated in terms of the interaction picture free fields. Secondly, boundary terms that only contain fields also do not contribute to field correlators. This is manifest in the Schwinger-Keldysh path-integral formalism, characterised by the apparent doubling of each degree of freedom $\varphi \to \varphi_{\pm}$ (see, e.g.~\cite{Chen:2017ryl}). In this representation, the expectation value of field operators at time $t$ is written as $\langle O(\varphi(t)) \rangle=\int {\cal D} \varphi_+ {\cal D} \varphi_- O(\varphi(t)) e^{i  \int^{t} dt' d \bx ({\cal L}[\varphi_+]-{\cal L}[\varphi_-])} \delta(\varphi_+(t)-\varphi_-(t))$, where the fields $\varphi_+$ and $\varphi_-$ are sewn at time $t$ (but not their time derivatives). Hence, the contribution in the weight $e^{i  \int^{t} dt' d \bx ({\cal L}[\varphi_+]-{\cal L}[\varphi_-])}$ of total derivative terms involving fields only is simply one, so that they can be discarded \cite{Arroja:2011yj,Burrage:2011hd}. Thirdly, the same argument does not hold in the case of total derivative terms involving time-derivative of fields, which contribute in general to correlators. However, cubic boundary terms like the one appearing explicitly in \eqref{forgotten}, which contain only the conjugate momenta of some field, here $\F$, only affect correlation functions involving that field (see, e.g.~\cite{Garcia-Saenz:2019njm}, appendix B). This is so because such boundary terms can be removed from the action by redefining the field that is involved (making appear terms proportional to the linear equations of motion which are innocuous, and terms beyond cubic order), a procedure that does not affect correlators of other fields. 

As a result, in order to compute the (one-loop) power spectrum of $\R$ as we wish, it is sufficient to consider only the bulk term \eqref{third_spectator}.
However, this is not yet in the best form for our purpose. Instead, it is convenient to rewrite the second term in \eqref{third_spectator} as 
\begin{align}\label{intparts}
-\epsilon a^2\F'\partial_i\F\partial_i\partial^{-2}\R' &= \epsilon a^2(\F''+2\mathcal{H}\F')\partial_i\F\partial_i\partial^{-2}\R +\epsilon a^2\F'\partial_{i}\F'\partial_i\partial^{-2}\R\\ \nonumber
&\!\!\!\!\!\!\!\!\!\!\!\!\!\!\!\!
=\epsilon a^2 \partial^2\F  \partial_i\F\partial_i\partial^{-2}\R -\frac{1}{2} \epsilon a^2(\F')^2\R     -\epsilon\frac{\delta \mathcal{L}^{(2)}}{\delta \F}\partial_i\F\partial_i\partial^{-2}\R \\ \nonumber
&\!\!\!\!\!\!\!\!\!\!\!\!\!\!\!\!
=\,-\epsilon a^2\partial^{-2}\left[\partial_i\partial^2\F \partial_i\F +(\partial^2\F)^2 \right]\R  -\frac{1}{2} \epsilon a^2(\F')^2\R   -\epsilon\frac{\delta \mathcal{L}^{(2)}}{\delta \F}\partial_i\F\partial_i\partial^{-2}\R.
\end{align}
On the first line we have integrated by parts the time derivative over $\R$, keeping only leading order terms in slow-roll (including the term proportional to $\epsilon'$ neglected here is straightforward and would not change our analysis). On the second line we rewrote the first term using the linear eom for $\F$ so that a term proportional to the linear eom appears. As discussed above, this term does not contribute to correlators, and we can discard it. 

Moreover, spatial integrations by parts have been performed, in particular between the first and the second line to remove a spatial derivative acting on $\F'$. 
We stress that these steps introduce boundary terms that contain no time derivative of $\R$, and hence, as explained above, they do not contribute to the equal-time $\R^n$ correlators. Lastly, note that the second term in the final expression cancels with the first term in \eqref{third_spectator}.

As a result, we have thus shown that for the purpose of this paper, it is sufficient to consider the following simple form of the Lagrangian, that is equivalent to the bulk Lagrangian (\ref{third_spectator}), and where $\R$ enters linearly and without time derivatives:
\begin{align} \label{L3bulk-final}
\mathcal {\tilde{L}}^{(3)}_{\mathrm{bulk}} = & \frac{1}{2}\epsilon a^2 (\partial_{i}\F)^2 \R -\epsilon a^2 \R \partial^{-2}\left[\partial_i\partial^2\F \partial_i\F +(\partial^2\F)^2 \right]  
\end{align}
(in the following, we omit the tilde to simplify the notation). Equipped with this, it is easy to compute the one-loop power spectrum, to which we now turn.

\subsection{In-in correlator}\label{subsec:inin}

According to what we discussed
the previous section, we need to compute~\footnote{In this expression we have added the suffix $I$ to indicate that the fields on the right hand side (both those explicitly written and those appearing in the interaction Hamiltonian) are the linear interaction-picture fields discussed in Section \ref{spectator}. For simplicity, this suffix is omitted in all other expressions of this paper.} 
\begin{equation}\label{twonested}
\langle\hat{\R}(\bk,\tau)\hat{\R}(\bk',\tau)\rangle=-\int_{\tauf}^{\tau}d\tau_{1}\int_{\tauf}^{\tau_{1}}d\tau_{2}\langle[\mathcal{H}^{(3)}_{I}(\tau_{1}),[\mathcal{H}^{(3)}_{I}(\tau_{2}),\hat{\R}_{I}(\bk,\tau)\hat{\R}_{I}(\bk',\tau)]]\rangle, 
\end{equation}
with the interaction Hamiltonian taking the form 
\be\label{H3}
\hat{\mathcal{H}}^{(3)} = \frac{\Mp^2}{(2\pi)^3}\int d \bk \,
2 a^2 \epsilon\, \hat{\R}(\bk,\tau) \hat{S}(-\bk,\tau) ,
\ee
where $\hat{S}$ is the Fourier transform of the functional derivative of (\ref{L3bulk-final}) with respect to $\R$ 
\be\label{source}
\hat{S} (\bk,\tau) 
= \int d \bx \, e^{-i\bk \bx} \, \hat{S}(\bx,\tau) 
\equiv \int d \bx \, e^{-i\bk \bx} \, 
\frac{1}{2 a^ 2 \epsilon \Mp^2}\frac{\delta\mathcal{L}_{\mathrm{bulk}}^{(3)}}{\delta \R}\,.
\ee
From the expression (\ref{L3bulk-final}) we obtain  
\be\label{source}
\hat{S} (\bx,\tau) \equiv \frac{1}{2 \epsilon a^2 \Mp^2}\frac{\delta\mathcal{L}_{\mathrm{bulk}}^{(3)}}{\delta \R} 
= \frac{1}{4 \Mp^2}\left( (\partial_i\Fh)^2 -\partial^{-2}\left[2(\partial^2\Fh)^2 + 2 \partial_i\partial^2\Fh\partial_i \Fh \right] \right) \equiv \hat{S}_{\F} (\bx,\tau) \,,
\ee
(where the suffix $\F$ in the last expression indicates the fields contributing to the source, to distinguish this from the cases considered in the following sections) 
and thus we see that $\hat{S}_{\F}$ commutes with $\hat{\R}$, as there is no quantum mixing at the linear level between the field $\hat{\R}$ and the field $\hat{\F}$ that constitutes the source in eq.~(\ref{source}). We also note that this expression does not depend on $\Fh'$. Symmetrizing over the last term and going to Fourier space we finally get
\begin{align}\label{source entropic}
\hat{S}_{\F} (\textbf{k},\tau) &= \frac{1}{\Mp^2} \int 
\frac{d \textbf{p}}{(2\pi)^3}
f_{\F}(\bp,\bk - \bp)\Fh(\bp,\tau) \Fh(\bk - \bp,\tau), 
\end{align}
with the kernel 
\begin{align}\label{fF}
    f_{\F}(\bp,\bq) &= -\frac{\bp\cdot\bq}{4} + \frac{1}{4|\bp+\bq|^2} \left[2 p^2 q^2 +\bp \cdot \bq (p^2+q^2)\right] \nonumber\\
    &= \frac{1}{2|\bp+\bq|^2} \left[ p^2 q^2 -(\bp \cdot \bq)^2\right] = \frac{p^ 2}{2}  \sin^2\theta,    
\end{align}
where the relation $\bq = \bk - \bp$ has been taken into account and where $\theta$ is defined as the angle between $\bk$ and $\bp$. We note that this kernel coincides with the one obtained in \cite{Fumagalli:2021mpc} for the production of gravitational waves from scalar excited states, up to a factor that accounts for the different spin of the produced field. This is due to the fact that in both cases the production is only due to gravitational interactions, which are universal. We see from the second to last expression that the kernel depends only on the magnitude of the three momenta, in a way that is symmetric on the two internal momenta. 

Inserting the expression (\ref{H3}) into the correlator 
(\ref{twonested}), taking into account that $\hat{\R}$ and $\hat{S}$ commute at all times, and using eq.~\eqref{commutator}, we find
\begin{align}  
\langle\hat{\R} (\bk,\tau)\hat{\R} (\bk',\tau)\rangle =& -i \int_{\tauf}^{\tau}d\tau_{2} \, 
g_{k'} \left( \tau ,\, \tau_2 \right) \int_{\tauf}^{\tau_{2}}d\tau_{1} 
\frac{\Mp^2}{(2\pi)^3}\int d \bp \, 2 \epsilon a^2 \left( \tau_1 \right) \nonumber\\
& \left\langle \left[ \hat{\R}(\bp,\tau_1) \hat{S}_{\F} (-\bp,\tau_1) ,\, \hat{\R} (\bk,\tau) \hat{S}_{\F} (\bk',\tau_2)  \right] \right\rangle 
+ \bk \leftrightarrow \bk' \,. 
\label{eq:first-step}
\end{align}
Similarly, and with obvious short-hand notations, the remaining commutator simply reads $[\hat{\R}_1 \hat{S}_1,\hat{\R} \hat{S}_2]=[\hat{\R}_1,\hat{\R}] \hat{S}_2 \hat{S}_1 +\hat{\R}_1 \hat{\R} [\hat{S}_1,\hat{S}_2]$. 
We are thus left with two terms: the first one involves the commutator of $\hat{\R}$, and the second one involves the commutator of the source $\hat{S}$. As computed in eq.~(\ref{source entropic}) the source is quadratic in the entropic field, so the first term is of ${\rm O } \left( \beta^4 \right)$. In the second term two entropic fields are commuted, and the resulting commutator is independent of the \bog coefficients, cf. the discussion after eq.~(\ref{commutator}). Therefore, the second term is of ${\rm O } \left( \beta^2 \right)$. In the limit of high excitation that we are considering, this second term can be disregarded with respect to the first one, and the source can be treated as a classical  function commuting with itself at different times. With this in mind, we obtain 
\be
\langle\hat{\R}(\bk,\tau)\hat{\R}(\bk',\tau))\rangle = \int_{\tauf}^{\tau}d\tau_{2}
g_{k'} \left( \tau ,\, \tau_2 \right) \int_{\tauf}^{\tau_{2}}d\tau_{1} g_k \left( \tau ,\, \tau_1 \right) 
\left\langle  \hat{S}_{\F}(\bk,\tau_1) \hat{S}_{\F}(\bk',\tau_2) \right\rangle 
+ \bk \leftrightarrow \bk' \;. 
\ee
As the source is commuting, this expression is symmetric under the exchange of the two internal times, so that we can replace the domain of integration as
\be
\int^\tau d\tau_2 \int^{\tau_2} d\tau_1 = \frac{1}{2}\int^ \tau d\tau_1 \int ^ \tau d\tau_2.
\ee
Noting that we can also invert the sign of the momentum in the two sources, and that the correlator forces $k' = k$, we then obtain 
\begin{align}\label{resultinin}
\langle\hat{\R}(\bk,\tau)\hat{\R}(\bk',\tau)\rangle &= \int^ \tau d\tau_1 \int^\tau\, d\tau_2 \, g_k(\tau,\tau_1)g_{k'}(\tau,\tau_2) \langle \hat{S}_{\F}(\bk,\tau_1) \hat{S}_{\F} (\bk',\tau_2) \rangle,
\end{align}
where we recall that this expression holds up to subdominant ${\rm O } \left( \beta^2 \right)$ corrections. 

\subsection{Leading term and classical limit in presence of large particle production} \label{subsec:classical}

Note that the leading order result written in \eqref{resultinin} is equivalent to what would have been obtained by solving the classical eom for the field $\R$, in terms of a Green function times a classically commuting source. 

In fact by varying the action $\mathcal{L}^{(2)} + \mathcal{L}^{(3)}$ one can write the equation of motion for $\Rh$ in Fourier space as
\be\label{Ooperator}
(\partial^2_\tau + 2 a H (1+\eta/2)\partial_\tau + k^ 2)\Rh(\bk,\tau) \equiv \mathcal{O}_k\, \hat{\R}({\bk},\tau) = \hat{\underline{S}}_{\F}(\bk,\tau),
\ee
where\footnote{As usual, boundary terms which do not contain first derivatives of the field $\R$ do not contribute to the eom of $\R$.} 
\be
\hat{\underline{S}}_{\F}(\bk,\tau)  \equiv \frac{1}{2 a^ 2 \epsilon\Mp^2}\frac{\delta(\mathcal{L}_{\mathrm{bulk}}^{(3)}+\mathcal{L}_{\mathrm{eom}}^{(3)})}{\delta \R}.
\label{Shatunder}
\ee
Solving the equation iteratively means that, at leading order (see below), one has to substitute the solution to the linear eom on the right hand side.
Since $\delta \mathcal{L}^{(2)}/\delta \F$ in $\mathcal{L}^{(3)}_{\mathrm{eom}}$ (eq. \eqref{eomterm}) is not touched by taking derivative with respect to $\R$, the corresponding contribution to the source term vanishes and the source term is simply given by $\hat{\underline{S}}_{\F}(\bk,\tau) = \hat{S}_{\F} (\bk,\tau) $ as defined from the bulk Lagrangian in \eqref{source}.
Thus the rescattered curvature fluctuation at second order can be written as
\be\label{classical}
\hat{\R}(\bk,\tau)= \int^\tau \di \tau' g_k(\tau,\tau')\hat{S}_{\F}(\bk,\tau') \,, 
\ee
where $g_k$ is the Green function already given in eq.~(\ref{RR-FF}) and associated with the operator appearing in the linear equation of motion, i.e. $\mathcal{O}_k$ in \eqref{Ooperator}.
Taking the two-point correlator of the operator \eqref{classical} then reproduces the result  \eqref{resultinin}.

Let us conclude this subsection expanding on the comment made after eq. (\ref{Shatunder}). With schematic notations, the $\R\, \F^2$ couplings in the action leads to the nonlinear equation of motion \eqref{Ooperator} for $\R$, but it also leads to a nonlinear source term for $\F$, i.e.~the latter obeys an equation of the type ${\cal O}_{\F} \F \sim \R \F$. Solving this coupled system iteratively (where higher-order terms involve larger number of linear fields), we can write $\R=\R^{(1)}+\R^{(2)} +\R^{(3)}+\ldots$ and similarly for $\F$, which gives two contributions to the power spectrum: $\langle \R^{(2)} \R^{(2)} \rangle $ and $\langle \R^{(1)}  \R^{(3)} + (1 \leftrightarrow 3) \rangle$. The first one corresponds to the leading-order effect described above, with the second one being negligible. Indeed, the first one schematically  involves $\langle (\F^{(1)})^4 \rangle$, while the second one involves $\langle (\F^{(1)})^2 (\R^{(1)})^2  \rangle $. These two contributions thus scale like $\beta^4$ and $\beta^2$ respectively, and the latter is negligible in the regime of high excitation that we consider. Naturally, this discussion at the level of the equation of motion finds its counterpart when using the in-in formalism, see the discussion after \eqref{eq:first-step}.

\subsection{One-loop power spectrum}
\label{subsec:1loop}

Let us evaluate the correlator \eqref{resultinin}, with the source $\hat{S}_{\F}$ given by \eqref{source entropic}.  Recalling eqs. \eqref{interactionfield}-\eqref{quantiz}, we obtain 
\begin{align}
\langle \hat{\R}(\bk,\tau)\hat{\R}(\bk',\tau) \rangle &= \frac{2 \delta^{(3)} \left( \bk+\bk'\right)}{\Mp^4} \int d\bp f_{\F}^2(\bp,\bk-\bp) \cdot\left|\int^\tau d\tau' g_k(\tau,\tau')\F(p,\tau')\F(|\bk-\bp|,\tau')\right|^2.
\end{align}
Introducing the dimensionless quantitites 
\be \label{xy}
x \equiv \frac{p}{k} \;\;,\;\; y \equiv \frac{|\bk-\bp|}{k} \;\;,\;\; z = k \tau' \;, 
\ee
we have the late time  ($\tau \to 0$) / super-horizon correlator
\begin{align}
\langle \hat{\R}(\bk,0)\hat{\R}(\bk',0) \rangle &= \frac{\pi H^4 \delta^{(3)} \left( \bk+\bk'\right)}{4 \Mp^4 k^3}  \int dx \int dy  \left( \frac{
\left( \left( x + y \right)^2-1\right)
\left( 1 - \left( x - y \right)^2\right)}{4 x y} \right)^2 \nonumber\\
&\!\!\!\!\!\!\!\! 
\left|\int_{z_{\rm out}}^0 dz \,i \frac{\U( 0 )\U^*( z )-\U^*( 0 )\U( z )}{2 z^2} \sum_{s_{1,2} = \pm } 
\alpha^{s_1}(k x) \U(s_1 x z)
\alpha^{s_2}(k y) \U(s_2 y z) \right\vert^2 \;, 
\end{align}
where we have exploited the fact that 
$\U^* (+x) = \U (-x)$, and defined 
\begin{equation}
\alpha^{+}(k) \equiv \alpha_{\F}(k) \;\;,\;\; 
\alpha^{-}(k) \equiv \beta_{\F}(k) \;. 
\end{equation}
We then use eq.~(\ref{defpower}) to go from the correlator to the power spectrum 
\begin{tcolorbox}
[colframe=white,arc=0pt,colback=lightgray!40]
\begin{equation}
\begin{aligned}
\mathcal{P}_{\R}^{\mathrm{1-loop}}(k) = \epsilon_0^2 \, \p_0^2 \, \int_0^{\infty} \di  x \int_{|1-x|}^{1+x} \di  y \,\mu^2_{\F}(x,y) \cdot \left|\sum_{{\rm s_{1,2}=\pm}} \alpha^{\rm s_{1}}(k x) \alpha^{\rm s_{2}}(k y) \cG(\mathrm{s_1}x,\mathrm{s_2}y,\zf) \right|^2 \;,
\label{eq:P-1-loop-final}
\end{aligned}
\end{equation}
\end{tcolorbox}
\noindent where 
\begin{equation}
\p_0 = \frac{H^2}{8 \pi^2 \Mp^2 \,\epsilon_0} 
\label{P0}
\end{equation}
is the power spectrum of the linear scalar perturbations
for the modes that left the horizon well before the generation of the excited state, and $\epsilon_0$ is the corresponding value of the slow-roll parameter. We also introduced 
\begin{eqnarray} 
&& \mu_{\F}(x,y) \equiv \frac{\left((x+y)^2-1\right) \left(1-(x-y)^2\right)}{4\, x y} \;, 
\label{muF}
\end{eqnarray} 
and
\begin{eqnarray} 
\label{master integral}
&& \cG(x,y,\zf) \equiv \int^{0}_{\zf}  \di z \, G(0,z)\U(x z)\U(y z) \;, 
\end{eqnarray} 
where $G$ is the rescaled Green function $G(z,z') \equiv k g_k (\tau,\tau')$, and $g_k$ is defined in eq.~\eqref{RR-FF}. 

This time integral  has an analytical representation (see eq.~4.34 in \cite{Fumagalli:2021mpc})
\begin{align}
\cG &= \mathcal{K}_1(x,y)-\mathcal{F}_1(x,y,\zf)-\mathcal{F}_1^*(-x,-y,\zf) \,,  
\label{cG_analytical}
\end{align}
where the first (respectively, second) term is obtained from the upper bound of integration $z=0$ (respectively lower bound $z_{\rm out}$). Evaluating the integral gives 
\begin{align} \label{K1F1}
\mathcal{K}_1(x,y) &= \frac{1-2x y-(x+y)^2}{(1-(x+y)^2)^2} \, , \\
\nonumber \mathcal{F}_1(x,y,\zf) &= \frac{e^{-i(1+x+y)\zf}}{2(1+x+y)^2} \; \Bigg[ \frac{i (1+x+y)^2}{\zf} -ixy(1+x+y) \zf \nonumber\\ 
& \hphantom{= \frac{e^{-i(1+x+y)\zf}}{2(1+x+y)^2} \bigg( } \, -x-y-(x+y)^2-xy(2+x+y) \Bigg] \,.\label{K1F1-2}
\end{align}
The time integration that we have just evaluated is at the root of the resonant effect which boosts the one-loop power spectrum. As we will see more explicitly in the next subsection, the enhancement takes place when the sourcing fields $\F \left( \bp \right)$ and $\F \left( \bq \right)$ (with $\bp + \bq = \bk$ being the momentum of the sourced mode) are deep inside the horizon at the transition, and the sourced modes $\R \left( \bk \right)$ that are resonantly enhanced the most have a wavelength comparable to the horizon at the transition. Therefore the enhancement is due to modes with $x \equiv p/k \simeq y \simeq q/k \gg 1$. In this regime, the integrand of \eqref{master integral} has potentially fast oscillating phases stemming from 
\begin{equation}
\U(s_1 x z) \, \U(s_2 y z) \propto {\rm e}^{-i \left( s_1 x + s_2 y \right) z} \;. 
\end{equation} 
For general large values of $x$ and $y$, these fast oscillating phases strongly reduce the magnitude of the integral. However, when $x=y$, this phase is absent for $s_1 = +1 ,\, s_2 = -1$ and for $s_1 = -1 ,\, s_2 = +1$. This is the mathematical reason for the resonant enhancement,\footnote{We can also see this at the level of the terms in eq.~\eqref{K1F1} and eq.~\eqref{K1F1-2}. For a sourced mode at the horizon scale, and a sourcing mode well inside the horizon, $z_{\rm out} = {\rm O } \left( 1 \right)$. When inserting in eq. (\ref{eq:P-1-loop-final}) these terms have to be evaluated with the replacements $x \to s_1 x = \pm x$ and $y \to s_2 y = \pm y$. For $y = x$ (namely, when the two signs $s_{1,2}$ coincide), the numerators of the two terms are of order $x^2$ and $x^3$ respectively. Instead, the two denominators are, respectively, of order $x^4$ and $x^2$, leading to, respectively, a ratio of order $x^{-2}$ and $x$ (in the last case, this growth with $x$ is accompanied by an $x$-dependent phase in eq.~\eqref{K1F1-2}, that suppresses the $x$ integration in eq. (\ref{eq:P-1-loop-final})). However, when $y = -x$ (namely, when the two signs $s_{1,2}$ are opposite), the two numerators are of order $x^2$, while the two denominators are of order one, leading to ratios of order $x^2$ that account for the resonant enhancement.} which physically corresponds to constructive interferences between positive and negative frequency modes, as first studied in \cite{Fumagalli:2021mpc} for scalar-scalar-tensor interactions. This is illustrated more clearly in the limit of peaked sources that we consider below, as the result of the integration in momentum space greatly simplifies in this limit.

\subsection{Analytical expression for peaked Bogoliubov coefficients}
\label{subsec:analytic-peaked-specfield}

In a wide range of models of interest the Bogoliubov coefficients $\alpha_k, \beta_k$ are sharply peaked. We model this case with a Gaussian profile
\begin{align}
\label{Gauss}
& \left\vert \alpha _k \right\vert = \left\vert \beta_k \right\vert \equiv \frac{\bs \, k_*^{1/2}}{\pi^{1/4} \, \sigma^{1/2}}  \, {\rm e}^{-\frac{\left( k - k_* \right)^2}{2 \sigma^2}} \;\;,\;\;\ \bs \gg 1 \;, \nonumber\\[1mm]
& \alpha_k = |\beta_k| e^{i \varphi_+} \;,\;
\beta_k = |\beta_k| e^{i \varphi_-} \,,
\end{align}
where $k_*$ and $\sigma$ are, respectively, the location and the width of the peak, and where the momentum dependence of the phases is understood. In these parametrizations, the two Bogoliubov coefficients are taken to have identical magnitude, which is true in the large occupation limit, see eq.~\eqref{quantization}. In the limit of very sharp peaks these relations reduce to Delta-functions profiles 
\be
\label{eq:alphak-betak-delta}
\lim_{\sigma / k_* \to 0} \left\vert \alpha_k \right\vert^2 = \lim_{\sigma / k_* \to 0} \left\vert \beta_k \right\vert^2 = \bs^2 \, k_* \, \delta \left( k - k_* \right)  = \bs^2 \, \delta \left( \ln \frac{k}{k_*} \right) \,. 
\ee 
For finite but sufficiently small width $\sigma$, the functions $\mu^2_{\F}$ and $\cG$ entering in eq.~(\ref{eq:P-1-loop-final}), as well as the phases of the two Bogoliubov coefficients, can be treated as constant near the peak, and we can approximate eq.~(\ref{eq:P-1-loop-final}) as 
\begin{align}
\mathcal{P}_{\R}^{\mathrm{1-loop}}(k) \simeq &  \epsilon_0^2 \, \p_0^2 \, 
\left( \frac{\bs \, k_*^{1/2}}{\pi^{1/4} \, \sigma^{1/2}} \right)^4 \, 
\mu^2_{\F} \left( \frac{k_*}{k} , \frac{k_*}{k} \right)
\left|\sum_{{\rm s_{1,2}=\pm}} 
{\rm e}^{i \left( \varphi_{s_1} +  \varphi_{s_2} \right)} \; \cG\left(\frac{\mathrm{s_1} \, k_*}{k},\frac{\mathrm{s_2} \, k_*}{k},\zf\right) \right|^2 \nonumber\\
& \times \int_0^{\infty} \di  x \int_{|1-x|}^{1+x} \di  y \, {\rm e}^{-\frac{\left( x k - k_* \right)^2 + \left( y k - k_* \right)^2}{\sigma^2}} \;. 
\end{align}
In terms of the two variables $s \equiv x+y$ and $d \equiv x-y$ we obtain two separate integrals in the domains $1 \leq s \leq \infty$ and $-1\leq d \leq 1$, which can be performed analytically to give 
\begin{align}
\label{eq:P-1-loop-finite-width}
\mathcal{P}_{\R}^{\mathrm{1-loop}}(k) &=  
\epsilon_0^2 \, \p_0^2 \, \frac{\bs^4}{\kappa^2}  \,\mu^2_{\F}\left( \kappa^{-1} ,\, \kappa^{-1} \right)
\, \left|\sum_{{\rm s_{1,2}=\pm}} 
{\rm e}^{i \left( \varphi_{s_1} +  \varphi_{s_2} \right)} \; \cG\left(\frac{\mathrm{s_1}}{\kappa},\frac{\mathrm{s_2}}{\kappa}, -\gamma \kappa \right) \right|^2 
\nonumber\\
& \times  {\rm Erf} \left( \frac{k}{\sqrt{2} \sigma} \right) 
\frac{1}{2} \left[ 1 - {\rm Erf} \left( \frac{k-2 k_*}{\sqrt{2} \sigma} \right) \right] \;, 
\end{align}
where ${\rm Erf}$ denotes the error function. In this expression we introduced the two parameters 
\begin{equation}
\kappa \equiv \frac{k}{k_*} \;\;,\;\; 
\gamma \equiv -k_* \tau_\textrm{out} = \frac{k_*}{k_{\rm{out}}}\, ,
\label{gamma}
\end{equation}
so that the last argument of $\cG$ indeed reads $\zf = \frac{k}{k_*} \, k_* \tau_{\rm out} = -\gamma \kappa$. The parameter $\gamma$ quantifies how subhorizon the excited mode $k_*$ was at the transition, namely at the time $\tauf = -1/k_{\rm{out}}$.

In the formal limit (\ref{eq:alphak-betak-delta}) of infinitely peaked Bogoliubov coefficients the result \eqref{eq:P-1-loop-finite-width} 
simplifies to 
\begin{align}
\mathcal{P}^{\mathrm{1-loop}}_{\R,\,\delta} \left( k \right) \simeq  
\epsilon_0^2\p_0^2\frac{\bs^4}{\kappa^2} \mu^2_{\F}\left(\kappa^{-1},\kappa^{-1}\right)\left|\sum_{{\rm s_{1,2}=\pm}} 
{\rm e}^{i \left( \varphi_{s_1} +  \varphi_{s_2} \right)} \; \cG\left(\frac{\mathrm{s_1}}{\kappa},\frac{\mathrm{s_2}}{\kappa},-\gamma \kappa\right) \right|^2 \Theta \bigg(2 - \kappa  \bigg)\,, 
\label{eq:P-1-loop-delta-schematic}
\end{align}
where $\Theta$ is the Heaviside theta function. 
This factor arises from $\frac{1}{2} \left[ 1 - {\rm Erf} \left( \frac{k-2 k_*}{\sqrt{2} \sigma} \right) \right]$ in the $\sigma \ll k_*$ limit, which is the regime we are considering. For this reason, from now on we make this replacement also for the case of finite width. The main difference between the finite width result \eqref{eq:P-1-loop-finite-width} and the approximation (\ref{eq:P-1-loop-delta-schematic}) is due to the other error function, which gives a different scaling at small momenta 
\be
\lim_{k \ll \sigma \ll k_*} 
\frac{\mathcal{P}_{\R}^{\mathrm{1-loop}} \left( k \right)}{\mathcal{P}^{\mathrm{1-loop}}_{\R,\,\delta} \left( k \right)} = \sqrt{\frac{2}{\pi}}\frac{k}{\sigma} \;,
\ee
improving over the analysis in \cite{Fumagalli:2021mpc}, bypassing the need to introduce a fudge factor.

\subsection{Limit of UV source}\label{subsec:UVsource}

The situation $\gamma \gg 1$ represents an interesting and physically motivated limit in which the excited entropic modes are deeply sub-horizon. This regime is, for instance, relevant for sharp-turn inflation models of type \cite{Palma:2020ejf, Fumagalli:2020adf, Fumagalli:2020nvq} with sufficiently large turn rate. As anticipated at the end of Subsection \ref{subsec:1loop}, the fact that the sourcing fields are deeply sub-horizon results in a resonant enhancement of $\R$ modes of the horizon size. This can be seen as an \textit{infrared rescattering}, since the produced modes of $\R$ have wavenumbers much smaller than that of the sourcing modes of $\F$. 

In the above discussion, we noted that, at the mathematical level, the enhancement is due to terms in which the first two arguments of the function $\cG$ entering in eq. (\ref{eq:P-1-loop-final}) have opposite signs, while the contributions in which the two arguments have equal sign can be disregarded. In the peaked case, in which the absolute values of the first two arguments are identical, the dominant contributions are \cite{Fumagalli:2021mpc} 
\be
\cG \left( \kappa^{-1},-\kappa^{-1},-\gamma \kappa \right) = \cG \left( - \kappa^{-1},\kappa^{-1},-\gamma \kappa \right) 
= \gamma^2 \bigg[- \frac{\sin(\gamma \kappa)}{\gamma \kappa} +\frac{2 \big(1- \cos(\gamma \kappa) \big)}{(\gamma \kappa)^2} +\frac{1}{\gamma^2} \bigg(1- \frac{\sin(\gamma \kappa)}{\gamma \kappa} \bigg) \bigg],
\label{cG-sincos}  
\ee
which has been organized in this way since the case of resonant enhancement corresponds to $\gamma \gg 1$ and $\gamma \kappa = - \zf = {\rm O } \left( 1 \right)$, as discussed above. We see that for $\gamma \gg 1$ the third term in this expression can be safely ignored, and we can write 
\begin{align}
\mathcal{P}_{\R}^{\mathrm{1-loop}}(k) & {\underset{\gamma \gg 1}{\simeq}} 4 \epsilon_0^2 \, \p_0^2 \, \bs^4 \frac{\gamma^2}{\kappa^4}  \,\mu^2_{\F}\left( \kappa^{-1} ,\, \kappa^{-1} \right)\left[\sin(\gamma \kappa)-\frac{2 \big(1- \cos(\gamma \kappa) \big)}{\gamma \kappa} \right]^2 \, {\rm Erf} \left( \frac{k}{\sqrt{2} \sigma} \right) \Theta \left( 2 - \kappa \right) \;. 
\label{eq:P-1-loop-delta-Bog-large-gamma-final} 
\end{align}
\noindent which is independent of the phases of $\alpha$ and $\beta$ introduced in eqs. (\ref{eq:alphak-betak-delta}).

\begin{figure}[t!]
\centerline{
\includegraphics[width=0.65\textwidth,angle=0]
{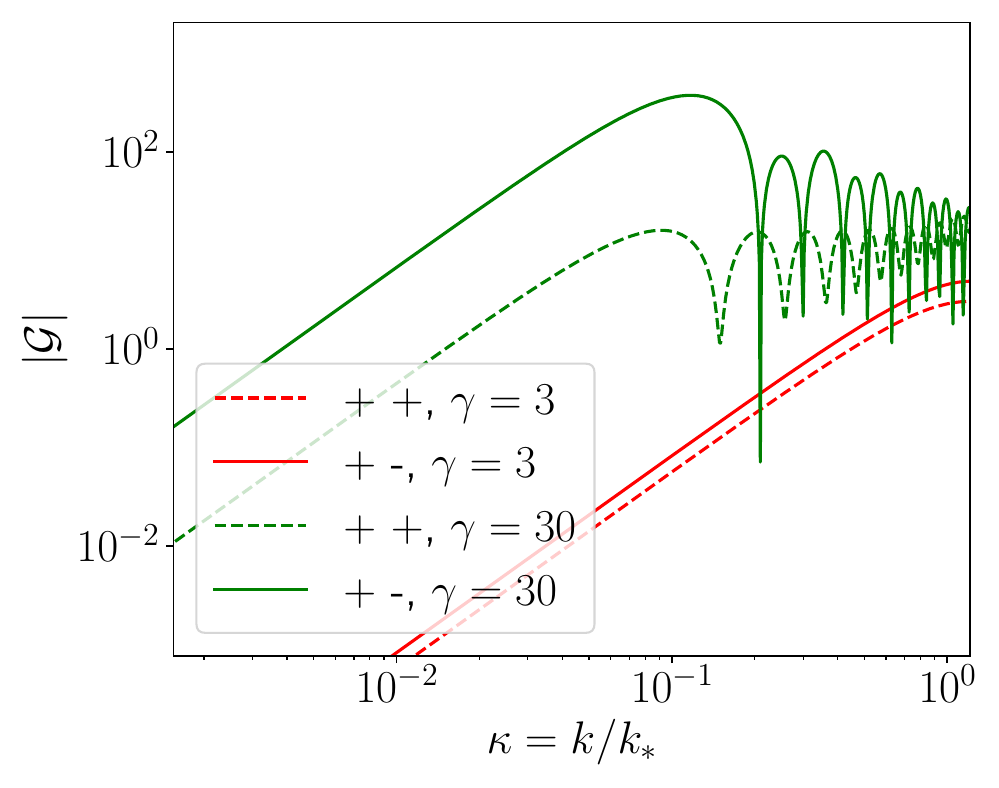}}
\caption{
Factor $\left\vert \cG \left( \pm \kappa^{-1} ,\, \pm \kappa^{-1} ,\, - \gamma \kappa \right) \right\vert$, evaluated through eq.~\eqref{cG_analytical}, as a function of $\kappa=k/k_*$ for two different values of $\gamma$. Solid lines (resp. dashed lines) refer to the $+ -$ (respectively $++$) choice for the signs of the first two arguments. We note the enhancement at $\gamma \gg 1$ for the case of opposite signs.}
\label{fig:cG_gamma_3_20}
\end{figure}

\begin{figure}[t!]
\centering
\includegraphics[width=0.78\columnwidth]{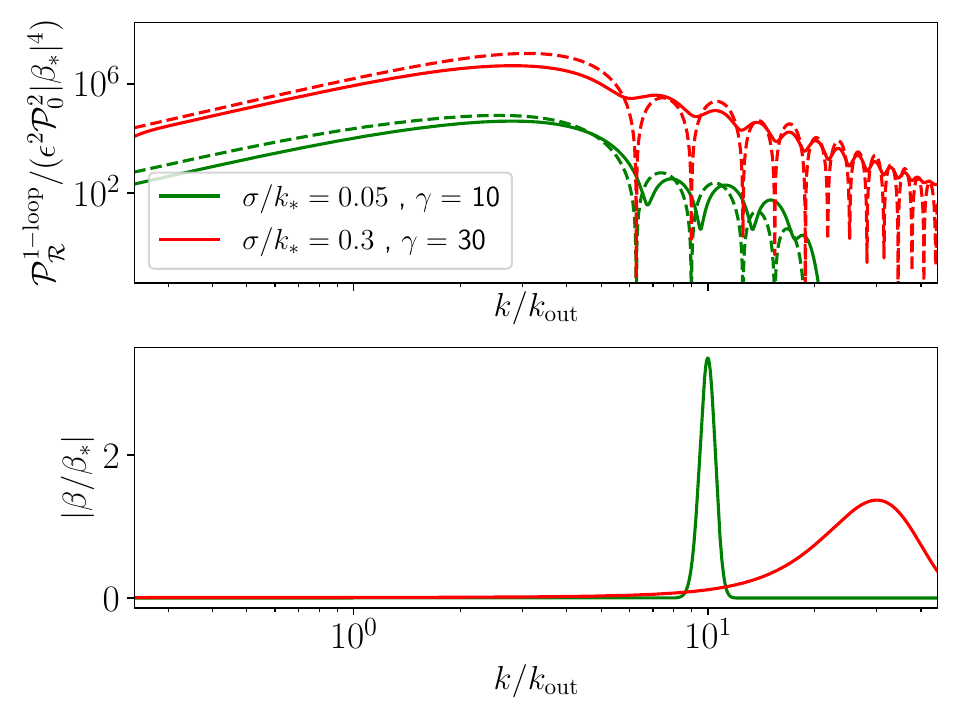}
\caption{\textit{Top panel:}
$\mathcal{P}_\R^{\textrm{1-loop}}(k)$ induced by an excited spectator field with large occupation number and peaked \bog coefficients. Solid lines represent $\mathcal{P}_\R^{\textrm{1-loop}}(k)$ calculated numerically using eq. \eqref{eq:P-1-loop-final} and a Gaussian profile for the \bog coefficients as in eq. \eqref{Gauss} and same phase for simplicity.
Dashed lines correspond to the analytical approximation of eq. \eqref{eq:P-1-loop-delta-Bog-large-gamma-final}. 
Examples are for two choices of $\gamma$ and $\sigma$ corresponding to the two limits in eqs. \eqref{twolimits1}-\eqref{twolimits2}, and the small difference in peak positions estimated in eq.~\eqref{twolimits2} is visible in the dashed curves.
\textit{Lower panel:} Shape of the \bog coefficients (normalized to their integral over $k$) indicating the amount of particles in the excited state.}
\label{figspectator3}
\end{figure}

The manifestation of the resonant enhancement is shown in Figure~\ref{fig:cG_gamma_3_20}. The figure shows the absolute value of the function $\cG$ as a function of the wavenumber $k$ of the sourced mode (more precisely, of the ratio $\kappa \equiv k / k_*$, where $k_*$ is the wavenumber at which the sourcing modes are peaked). Let us denote $\cG \left( s_1 ,\, s_2 \right) \equiv \cG \left( \frac{s_1}{\kappa} ,\, \frac{s_2}{\kappa} ,\, - \gamma \kappa \right)$ in the present discussion, with $s_{1,2} = \pm 1$, keeping in mind that this function is symmetric in its first two arguments. The red curves in the panel show the values of $\left\vert \cG \left( + ,\, + \right) \right\vert$ and of $\left\vert \cG \left( + ,\, - \right) \right\vert$ for $\gamma = 3$, namely for a case in which the sourcing field is only slightly inside the horizon when it is excited. We see that in this case, the two terms provide a comparable and non-enhanced contribution. The green curves show instead the values assumed by the same two functions for $\gamma = 30$, namely for the case in which the sourcing field is well inside the horizon when it is excited. We see that indeed $\left\vert \cG \left( + ,\, - \right) \right\vert \gg \left\vert \cG \left( + ,\, + \right) \right\vert$ in this case, and that the former contribution is enhanced. 
In Figure \ref{figspectator3} we show the one-loop power spectrum of $\R$ from eq.~(\ref{eq:P-1-loop-delta-Bog-large-gamma-final}) (with the combination $\frac{1}{2} \left[ 1 - {\rm Erf} \left( \frac{k-2 k_*}{\sqrt{2} \sigma} \right) \right]$ replacing the Heaviside theta function). The curves shown exhibit a maximum followed by a series of oscillations. This can be understood by noting that in this regime ($\kappa = \frac{{\rm O } \left( 1 \right)}{\gamma} \ll 1$) the factor $\mu_{\F}\left( \kappa^{-1} ,\, \kappa^{-1} \right)$ is approximately $1$, and by considering the momentum dependence in the two limiting cases of small and large arguments of the error function 
\begin{align}\label{twolimits1}
\mathcal{P}_{\R}^{\mathrm{1-loop}}(k) & {\underset{\gamma \gg 1}{\simeq}} 4 \epsilon_0^2 \, \p_0^2 \, \bs^4  \Theta \left( 2 - \kappa \right) \times \left\{ \begin{array}{l} 
\sqrt{\frac{2}{\pi}} \, \frac{k_*}{\sigma}
\frac{\gamma^5}{\left( \gamma \kappa \right)^3}  \,\left[\sin(\gamma \kappa)-\frac{2 \big(1- \cos(\gamma \kappa) \big)}{\gamma \kappa} \right]^2 \;\;,\;\; k \ll \sigma \\ 
\frac{\gamma^6}{\left( \gamma \kappa \right)^4}  \,\left[\sin(\gamma \kappa)-\frac{2 \big(1- \cos(\gamma \kappa) \big)}{\gamma \kappa} \right]^2 \;\;,\;\; k \gg \sigma
\end{array} \right. \;. 
\end{align}
Maximizing numerically the expressions in the two lines, we can find an approximate relation for the maximum value obtained by the one-loop correction
\be \label{twolimits2}
\mathcal{P}_{\R}^{\mathrm{1-loop}} \vert_{\rm max} \simeq \left\{ \begin{array}{l}
0.17 \, \epsilon_0^2 \, \p_0^2 \, \bs^4  \frac{k_* \, \gamma^5}{\sigma} \,  \;\;,\;\; 
{\rm at } \; \kappa \simeq \frac{3.1}{\gamma} \;\;,\;\; {\rm for} \;\; \frac{1}{\gamma} \ll \frac{\sigma}{k_*} \ll 1  \\ 
0.07 \, \epsilon_0^2 \, \p_0^2 \, \bs^4  \gamma^6 \,  \;\;,\;\; 
{\rm at } \; \kappa \simeq \frac{2.6}{\gamma} \;\;,\;\; {\rm for} \;\; \frac{\sigma}{k_*} \ll \frac{1}{\gamma}  \ll 1  
\end{array} \right. \;. 
\ee 
It is important to keep in mind that $\gamma$ cannot be arbitrarily large: avoiding backreaction on the background and strong coupling of fluctuations impose bounds on how much inside the horizon the excited modes can be, see e.g.~\cite{Fumagalli:2020nvq,Inomata:2021zel, Inomata:2021tpx, Fumagalli:2021mpc,Inomata:2022yte} for recent analyses in the general context of enhanced fluctuations. While these limits need to be taken into account in any full analysis of any specific model, several considerations prompt us not to apply these limits sharply in the examples we have considered. Firstly, we stress that such bounds are not theoretical requirements on the models \textit{per se}, but reflect limitations of the existing analytical results for the linear modes. Secondly, the existing bounds are based on estimates of these effects, and, to be applied sharply, they would need to be corroborated by more detailed computations and/or lattice simultations. Finally, they arise from the direct couplings that are responsible for the excitations of the linear modes, and, for this reason, they carry some model dependency (while our computation of the infrared rescattering are based on the universally present gravitational interactions).

\section{Rescattering in single-field inflation \textit{alias} self-scattering}\label{sec:single}

\subsection{General formula}
\label{subsec:general-self}
Let us now consider the situation where the excited scalar field is the curvature perturbation itself.
As a starting point, we consider the standard single-field cubic action in the $\delta\phi=0$ gauge \cite{Maldacena:2002vr}, as, for instance, recast in \cite{Burrage:2011hd,Garcia-Saenz:2019njm}.
There, not only the explicit size of the self-interactions $\sim O(\epsilon,\eta)^2$ in the bulk Lagrangian is manifest but also boundary terms are adjusted, in that context, to give negligible contribution on super-horizon scales to equal-time correlators. Similarly to the previous section, in the limit of large occupation number, we consider the 1-loop contribution to the power spectrum coming from the nonlinear equation of motion as derived from the cubic third-order bulk Lagrangian. The latter reads \cite{Burrage:2011hd,Garcia-Saenz:2019njm}\footnote{Although it is reasonable to assume that in the limit of large occupation number the Green function method provides the leading order result, a full proof from an in-in calculation in this context, as provided in the previous section for the excited isocurvature field, goes beyond the scope of this work.}

\begin{align}\label{third_zeta}
\mathcal{L}^{(3)} =\,\, & \Mp^2 a^2 \left[   \epsilon(\epsilon - \eta) \R(\R')^2 +  \epsilon(\epsilon + \eta)   \R(\partial_i \R)^2   + \frac{1}{a^2} \left( \frac{\epsilon}{2} - 2 \right) \partial_i \R \partial_i \chi \partial^2 \chi + \frac{\epsilon}{4 a^2} \partial^2 \R (\partial_i \chi )^2 
\right] \,,
\end{align}
where $\partial^2\chi = \epsilon a \R'$ (not important and not used below, so in the main text).
Disregarding terms that are of third order in the slow-roll parameter $\epsilon$, we rewrite eq. (\ref{third_zeta}) as 
\begin{align} 
\mathcal{L}^{(3)} \simeq  \Mp^2 a^2 \left[ \epsilon(\epsilon - \eta) \R(\R')^2 +  \epsilon(\epsilon + \eta)   \R(\partial_i \R)^2   - 2 \epsilon^2 \, \partial_i \R \, \partial_i \left( \partial^{-2} \R' \right) \R'
\right]. 
\label{Lag-RRR}
\end{align}
Extremizing the action obtained from this contribution and from the quadratic Lagrangian (\ref{quadraticR}), we obtain the equation of motion 
\bea 
\R''  +2 {\cal H}  \R'  -  \partial^2 \R 
&=&  \frac{\epsilon}{2} \Bigg\{ 4 \partial_i \R' \, \partial_i \left( \partial^{-2} \R' \right)  +  \left( \partial_i \R \right)^2  - 4 \R \partial^2 \R   - 2 \partial^{-2} \partial_i \left( \partial_i \R  \, \partial^2 \R  \right)   \Bigg\} \nonumber\\ 
&+& \frac{\eta}{2}  \left[  \R^{'2}  -  \left( \partial_i \R \right)^2     \right] \;. 
\eea 
In the extremization we disregarded derivatives of the slow-roll parameters, as they give terms suppressed by higher order in slow-roll. We also used the linear equation of motion (namely, the left-hand side of this equation equated to zero) to remove terms proportional to $\R''$ in the right-hand side. Similar to the discussion in section \ref{subsec:classical} for the excited entropic field, and with the same schematic notations, this corresponds to the leading contribution $\langle \R^{(2)} \R^{(2)} \rangle $, while neglecting $\langle \R^{(1)}  \R^{(3)} + (1 \leftrightarrow 3) \rangle$ (note that $\R^{(3)}$ includes also the effect of a single quartic interaction). Both involve $\langle (\R^{(1)})^4 \rangle$ and have similar contributions at the scales that are excited, at which the one-loop correction to the power spectrum should anyway be small compared to the tree-level one to ensure perturbative control. However, \textit{for the IR scales of interest in this work}, the first one is dominant, simply as it involves four excited modes, while the second contribution involves only two of them. In other words, only $\langle \R^{(2)} \R^{(2)} \rangle$ describes the IR rescattering of UV modes studied in this work.

Using the same notation as in Subsection \ref{subsec:classical}, we can write the equation of motion in Fourier space in the form 
\be
\left[ \partial^2_\tau + 2 a H \partial_\tau + k^2 \right] \Rh \left( \bk,\tau \right) = \hat{S}_{\R} (\bk,\tau) \;,
\label{eqR-self}
\ee
where now
\be
\label{Source zeta}
\hat{S}_{\R} (\bk,\tau) = 2 \epsilon \int \frac{d \bp}{\left( 2 \pi \right)^3} 
\Bigg[ f_{\R'} \left( \bp ,\, \bk - \bp \right) \R' \left( \bp ,\, \tau \right)  \R' \left( \bk - \bp ,\, \tau \right)  + f_{\R} \left( \bp ,\, \bk - \bp \right) \R \left( \bp ,\, \tau \right)  \R \left( \bk - \bp ,\, \tau \right) \Bigg] \;, 
\ee 
with 
\begin{align}\label{fRp}
f_{\R'} \left( \bp ,\, \bq \right) =&  \left( \frac{1}{p^2} + \frac{1}{q^2} \right) \frac{\bp \cdot \bq}{2} + \frac{\eta}{4 \epsilon} \;, \\ 
\label{fR}
f_{\R} \left( \bp ,\, \bq \right) =& 
\left[   \frac{p^2 +  q^2}{2} - \frac{\bp \cdot \bq }{4}+ \frac{2  p^2 q^2 + \left( p^2 + q^2 \right) \bp \cdot \bq }{4 \left\vert \bp+\bq \right\vert^2}   \right] + \frac{\eta}{4 \epsilon} \bp \cdot \bq \;. 
\end{align}
The solution of eq.~(\ref{eqR-self}) is formally given by eq.~(\ref{classical}) with the change in the source. Evaluating the new source correlator, the two-point function of the solution reads 
\bea
&& \!\!\!\!\!\!\!\!  \!\!\!\!\!\!\!\!  
\left\langle \hat{\R}(\bk,\tau) \hat{\R}(\bk',\tau) \right\rangle = 
\delta ^{(3)}\left( \bk+ \bk' \right) 8 \epsilon^2 \int d \bp 
\Bigg\vert \int^\tau d \tau' g_k \left( \tau ,\, \tau' \right) 
\Bigg[ 
f_{\R'} \left( \bp ,\, \bk - \bp \right) \R'(p,\tau') \R'(\vert \bk - \bp \vert,\tau') \nonumber\\ 
&& \quad\quad\quad\quad \quad\quad\quad\quad \quad\quad\quad\quad  \quad\quad\quad\quad 
\quad\quad\quad\quad 
\quad\quad
+ f_{\R} \left( \bp ,\, \bk - \bp \right) \R(p,\tau') \R(\vert \bk - \bp \vert,\tau') 
\Bigg] 
\Bigg\vert^2 \,. \nonumber\\ 
\eea 
From this expression, the computation of the one-loop power spectrum follows the same steps as in Subsection \ref{subsec:1loop}. To follow similar notations, we denote the Bogoliubov coefficients of $\R$ with a  compact notation as $\alpha^+_{\R} \equiv \alpha_{\R}$ and $\alpha^-_{\R} \equiv \beta_{\R}$. We obtain 
\begin{tcolorbox}
[colframe=white,arc=0pt,colback=lightgray!40]
\begin{equation}
\begin{aligned}
\mathcal{P}_{\R}^{\mathrm{1-loop}}(k) \simeq\,&\,  \epsilon_0^2 \p_0^2 \, \int_0^\infty d x \int_{\vert x-1 \vert}^{x+1} d y \; \Bigg\vert  \sum_{s_{1,2} = \pm}  \alpha_{\R}^{s_1} \left( k x \right)   \alpha_{\R}^{s_2} \left( k y \right) \\ 
& \quad
\times \left[  
\mu_{\R'} \left( x ,\, y \right) \;  \bar{\cG}( s_1 x, s_2 y,\zf) + 
\mu_{\R} \left( x ,\, y \right)  {\cG}( s_1 x, s_2 y,\zf)  \right] \Bigg\vert^2 , 
\label{eq:P-1-loop-final-self} 
\end{aligned}
\end{equation}
\end{tcolorbox}
\noindent where we introduced the functions 
\bea
\mu_{\R'}(x,y) &\equiv& \frac{2 f_{\R'}(x,y)}{x y } = \frac{1}{x y} \left[ 
\frac{\left(1-x^2-y^2\right)\left(x^2+y^2\right)}{2 x^2 y^2}  + \frac{\eta}{2 \epsilon} \right]  \;, \nonumber\\ 
\mu_{\R}(x,y) &\equiv& \frac{2 f_{\R}(x,y)}{k^2 x y}  = \frac{1}{x y} \left[ x^2+y^2+x^2 y^2 - \frac{\left( 1 - x^2 - y^2 \right)^2}{4} + \frac{\eta}{4 \epsilon} \left( 1 - x^2 - y^2 \right) \right] \, , \nonumber\\ 
\label{muR-Rp}
\eea
as well as the second master integral: 
\be
\label{master integral 2}
\bar{\cG}(x,y,\zf) \equiv x^ 2 y ^ 2 \int^{0}_{\zf}  \di z\, G(0,z) z^2 e^{-i( x + y)z},
\ee
which adds up to the master integral $\cG$, introduced in eq.~(\ref{master integral}), that was already present in the isocurvature case. Analogously to what we found in that case, see eq.~(\ref{cG_analytical}), this integral has the analytical representation 
\begin{align}
\bar{\cG} &= \mathcal{K}_2(x,y)-\mathcal{F}_2(x,y,\zf)-\mathcal{F}_2^*(-x,-y,\zf) \;, 
\label{barcG_analytical}
\end{align}
where the first (respectively, second) term is obtained from the upper bound of integration $z=0$ (respectively lower bound $z_{\rm out}$). Evaluating the integral gives 
\begin{align} \label{K2F2}
\mathcal{K}_2(x,y) &= \frac{2 x^2 y^2}{\left( 1-\left( x+y \right)^2 \right)^2} \, , \\
\mathcal{F}_2(x,y,\zf) &= \frac{e^{-i(1+x+y)\zf}}{2 \left( 1+x+y \right)^2} \, x^2 y^2 \, \Bigg[ i \left( 1 + x + y \right) \zf + 2 + x + y \Bigg]  \,.
\end{align}
As for the first master integral, this term is also enhanced for $x ,\, y \gg 1$ and for $s_1 = - s_2$, for the same physical reasons already discussed at the end of Subsection \ref{subsec:1loop}. Specifically, an enhancement of the power of $\R$ at some sub-horizon scale $k_*$ results in a subsequent resonant growth of the modes of $\R$ at scales $k_{\rm sourced} \equiv \kappa \, k_* \ll k_*$. As $x$ and $y$ are the ratios between the sourcing and the sourced momenta, this corresponds to $x \simeq y \simeq \kappa^{-1} \gg 1$. The modes $k_{\rm sourced}$ for which the resonant effect is greatest are at the horizon scales. For these modes, $\kappa \simeq \gamma^{-1}$, where $\gamma \gg 1$ quantifies how deep inside the horizon the sourcing mode $k_*$ is.~\footnote{This discussion summarizes the one of Subsection \ref{subsec:analytic-peaked-specfield} for the entropic case, as the physical reason for the enhancement is identical, and the mathematical expressions are analogous. See eqs. (\ref{gamma}) for the precise definitions of $\kappa$ and $\gamma$.} Eq. (\ref{cG-sincos}) manifestly shows how this enhancement mathematically arises from the first master integral $\cG$. In complete analogy, the second master integral evaluates to 
\be
\bar{\cG} \left( \kappa^{-1},-\kappa^{-1},-\gamma \kappa \right) = \bar{\cG} \left( - \kappa^{-1},\kappa^{-1},-\gamma \kappa \right) 
= \frac{\gamma^4}{\left( \gamma \kappa \right)^2} \bigg[- \frac{\sin(\gamma \kappa)}{\gamma \kappa} +\frac{2 \big(1- \cos(\gamma \kappa) \big)}{(\gamma \kappa)^2} \bigg] \;,
\label{barcG-sincos} 
\ee
which also manifestly shows the enhancement taking place for $\gamma \gg 1$ and $\kappa \, \gamma = {\rm O } \left( 1 \right)$. Naturally, the analytical reasoning in Section \ref{subsec:analytic-peaked-specfield} for peaked Bogoliubov coefficients also holds here, with the same parametric dependence of the amplitude of the IR peak, as $\mu_{{\cal R}'}(\kappa^{-1},\kappa^{-1})\simeq - \kappa^2(2-\eta/(2 \epsilon))$ and $\mu_{{\cal R}}(\kappa^{-1},\kappa^{-1}) \simeq 3-\eta/(2 \epsilon)$ for $\kappa \ll 1$, compensating in eq. (\ref{eq:P-1-loop-final-self}) the different scalings in $\kappa$ of ${\cG}$ and $\bar{\cG}$, see eqs.~\eqref{cG-sincos} and \eqref{barcG-sincos}.

\subsection{Example: phase of large $\eta$}\label{subsec:examplesingle}

Let us consider an explicit example where an excited state is dynamically generated in a single-field context. The set-up consists of a transient phase characterized by a strong departure from a slow-roll dynamics, which can excite quanta associated with the comoving curvature perturbation $\R$.

In particular, we consider a phase where the magnitude of the second slow-roll parameter $\eta \equiv \dot{\epsilon}/(\epsilon H)$, where $\epsilon \equiv - \dot{H}/H^2$, becomes much larger than one in between two standard slow-roll phases in which  $\epsilon,\vert \eta \vert \ll 1$.
This can happen, for instance, in presence of a step-like feature in the single-field inflaton potential \cite{Starobinsky:1992ts} such as an upward or downward step \cite{Inomata:2021tpx}.
More broadly, these types of dynamics \cite{Motohashi:2014ppa,Inoue:2001zt,Tzirakis:2007bf,Inomata:2021tpx,Tasinato:2023ukp} may be seen as a generalization of the so called ultra-slow-roll phase \cite{Tsamis:2003px,Kinney:2005vj,Namjoo:2012aa}, as the feature may lead to a phase in which the speed of the inflaton is greatly decreased. The jump in $\eta$ can be modelled as %
\begin{equation}\label{tophat}
\eta = \eta_c \, \theta \left( \tau - \tau_{\rm in} \right) \theta \left( \tauf - \tau \right),
\end{equation}
namely $\eta$ is non-vanishing only for times between $\tau_{\rm in}$ and $\tauf$, and it is constant in this interval.~\footnote{In Ref. \cite{Fumagalli:2023hpa} it has been shown that, \textit{irrespectively} of the sharpness of the transition, these transient non-slow-roll dynamics do not lead to a common rescaling of the power spectrum in the asymptotic IR.} It is immediate to extend this including also a value $\left\vert \eta \right\vert \ll 1$ outside this region, as this modifies the results by standard slow-roll corrections. 

As long as $\epsilon$ remains small, we can approximate $a \simeq - \frac{1}{H \tau}$ so that, from the definition of $\eta$, we obtain $\epsilon' = - \frac{\eta_c \, \epsilon}{\tau}$ in the time interval of the feature, leading to the continuous evolution for $\epsilon$:
\begin{eqnarray}
\epsilon &=& \epsilon_{\rm in} \;\;\;,\;\;\;\;\;\;\;\;\;\;\;\;\;\;\; \tau \leq \tau_{\rm in} \;\;, \nonumber\\ 
\epsilon &=& \epsilon_{\rm in} \left( \frac{\tau_{\rm in}}{\tau} \right)^{\eta_c} \;\;\;,\;\;\;\;\;\;\; \tau_{\rm in} \leq \tau \leq \tauf \;\;, \nonumber\\
\epsilon &=& \epsilon_{\rm in} \left( \frac{\tau_{\rm in}}{\tauf} \right)^{\eta_c} \equiv \epsilon_{\rm out}  \;\;\;,\;\;\; \tau \geq \tauf  \;. 
\label{eps-jumpeta}
\end{eqnarray} 
The duration $\Delta N$ of the non-slow-roll phase in \textit{e}-folds is related to these parameters by $\log \left( \frac{\epsilon_{\rm out}}{\epsilon_{\rm in}} \right) = \eta_c \, \log \left( \frac{\tau_{\rm in}}{\tauf} \right) = \eta_c \, \Delta N$. The transition (\ref{eps-jumpeta}) excites the linear mode function $\R$ to a form described by the general expression (\ref{excited1}):
\be\label{solsingle}
\R_k(\tau>\tauf) =\frac{H}{2 M_p \sqrt{\epsilon_{\rm out} k^3}} \left(\alpha_k \, \U(k\tau) + \beta_k \, \U^*(k\tau)\right).
\ee
The linear power spectrum on super-horizon scales is then given by
\be
\mathcal{P}_\R \left( k \right) = 
\frac{H^2}{8 \pi^2 \Mp^2 \,\epsilon_{\rm out}} \cdot \left\vert \alpha_k + \beta_k \right\vert^2 
 \;, 
\label{PR-excited}
\ee
where we note that the first factor is the linear power spectrum of a non excited state generated well after the phase of large $\eta$. 

Our goal is to compute at one-loop the self-scattering of this excited field using the formalism developed in the previous subsection. Before doing so, we review and briefly discuss the computation of 
(\ref{solsingle}) from the transition (\ref{eps-jumpeta}). The evolution (\ref{eps-jumpeta}) is characterized by three phases of $\epsilon \ll 1$ and of constant $\eta$. The linear solution for $\R_k$ in these phases is 
\be
{\R}_{k,i} = C_i  \, x^{\nu_i} H_{\nu_i}^{(1)} \left( x \right) + D_i \, x^{\nu_i} H_{\nu_i}^{(2)} \left( x \right) \;\;,\;\; x \equiv - k \tau \;\;,\;\; 
\nu_i \equiv \frac{3+\eta_i}{2} \;\;,\;\; i=``{\rm in }",\,``{\rm feature}",\, ``{\rm out}" \,, 
\label{R123-sol}
\ee
where $C_i$ and $D_i$ are integration constants, while the suffix $i$ refers to the three phases. The evolution (\ref{eps-jumpeta}) corresponds to 
\be 
\nu_{\rm in} = \nu_{\rm out} = \frac{3}{2} \;\;,\;\; \nu_{\rm feature} = \frac{3+\eta_c}{2} \;. 
\ee 
In the first time interval we choose the solution in the Bunch-Davies vacuum, namely $\alpha = 1$ and $\beta = 0$ in eq. (\ref{excited1}): 
\be
\lim_{- k \tau \to \infty} \R_{k,{\rm in}} = \frac{-i}{\sqrt{2 \epsilon_{\rm in}} M_p a} \, \frac{{\rm e}^{-i k \tau}}{\sqrt{2 k}} \;\;\; \Rightarrow \;\;\; 
C_{\rm in} = i \, \sqrt{\frac{\pi}{8 \epsilon_{\rm in} k^3}} \, \frac{H}{M_p} \;\;\;,\;\;\; D_{\rm in} = 0 \;. 
\end{equation}
We then impose continuity of $\R_k$ and of its time derivative at the two transition times $\tau_{\rm in}$ and $\tau_{\rm out}$, obtaining the expressions for the integration constants $C_{\rm out}$ and $D_{\rm out}$ after the two transitions. Finally, by identifying the expressions (\ref{solsingle}) and (\ref{R123-sol}) after the second transition, we obtain the Bogoliubov coefficients 
\begin{align}
\alpha_k &=  \sqrt{\frac{\epsilon_{\rm out}}{\epsilon_{\rm in}}} \, \frac{C_{\rm out}}{C_{\rm in}} =  \sqrt{\frac{\epsilon_{\rm out}}{\epsilon_{\rm in}}} \, 
\frac{\pi}{8} \, {\rm e}^{i \left( x_{\rm in} - x_{\rm out} \right)} {x_{\rm in}}^{1-\nu_c}  {x_{\rm out}}^{\nu_c-2} \nonumber\\ 
&\quad\quad\times \Bigg\{ \left[ \left( 1 + i x_{\rm out} \right) H_{\nu_c-1}^{(1)} \left( x_{\rm out} \right) - x_{\rm out} H_{\nu_c}^{(1)} \left( x_{\rm out} \right) \right] 
\left[ \left( 1 - i x_{\rm in} \right) H_{\nu_c-1}^{(2)} \left( x_{\rm in} \right) - x_{\rm in} H_{\nu_c}^{(2)} \left( x_{\rm in} \right) \right] \nonumber\\ 
&\quad\quad -  \left[ \left( 1 + i x_{\rm out} \right) H_{\nu_c-1}^{(2)} \left( x_{\rm out} \right) - x_{\rm out} H_{\nu_c}^{(2)} \left( x_{\rm out} \right) \right] 
\left[ \left( 1 - i x_{\rm in} \right) H_{\nu_c-1}^{(1)} \left( x_{\rm in} \right) - x_{\rm in} H_{\nu_c}^{(1)} \left( x_{\rm in} \right) \right] \Bigg\} , 
\label{alpha-jump}
\end{align}
and  
\begin{align}
\beta_k &= - \sqrt{\frac{\epsilon_{\rm out}}{\epsilon_{\rm in}}} \, \frac{D_{\rm out}}{C_{\rm in}}  = 
- \sqrt{\frac{\epsilon_{\rm out}}{\epsilon_{\rm in}}} \, 
\frac{\pi}{8} \, {\rm e}^{i \left( x_{\rm in} + x_{\rm out} \right)} {x_{\rm in}}^{1-\nu_c}  {x_{\rm out}}^{\nu_c-2} \nonumber\\ 
&\quad\quad\times \Bigg\{ \left[ \left( 1 - i x_{\rm out} \right) H_{\nu_c-1}^{(1)} \left( x_{\rm out} \right) - x_{\rm out} H_{\nu_c}^{(1)} \left( x_{\rm out} \right) \right] 
\left[ \left( 1 - i x_{\rm in} \right) H_{\nu_c-1}^{(2)} \left( x_{\rm in} \right) - x_{\rm in} H_{\nu_c}^{(2)} \left( x_{\rm in} \right) \right] \nonumber\\ 
&\quad\quad - \quad \left[ \left( 1 - i x_{\rm out} \right) H_{\nu_c-1}^{(2)} \left( x_{\rm out} \right) - x_{\rm out} H_{\nu_c}^{(2)} \left( x_{\rm out} \right) \right] 
\left[ \left( 1 - i x_{\rm in} \right) H_{\nu_c-1}^{(1)} \left( x_{\rm in} \right) - x_{\rm in} H_{\nu_c}^{(1)} \left( x_{\rm in} \right) \right] \Bigg\} \;,
\label{beta-jump}
\end{align} 
where $x_{\rm in} \equiv -k\tau_{\rm in}$ and $x_{\rm out} \equiv -k\tauf$. 

\begin{figure}[]
\centering
\includegraphics[width=0.47\columnwidth]{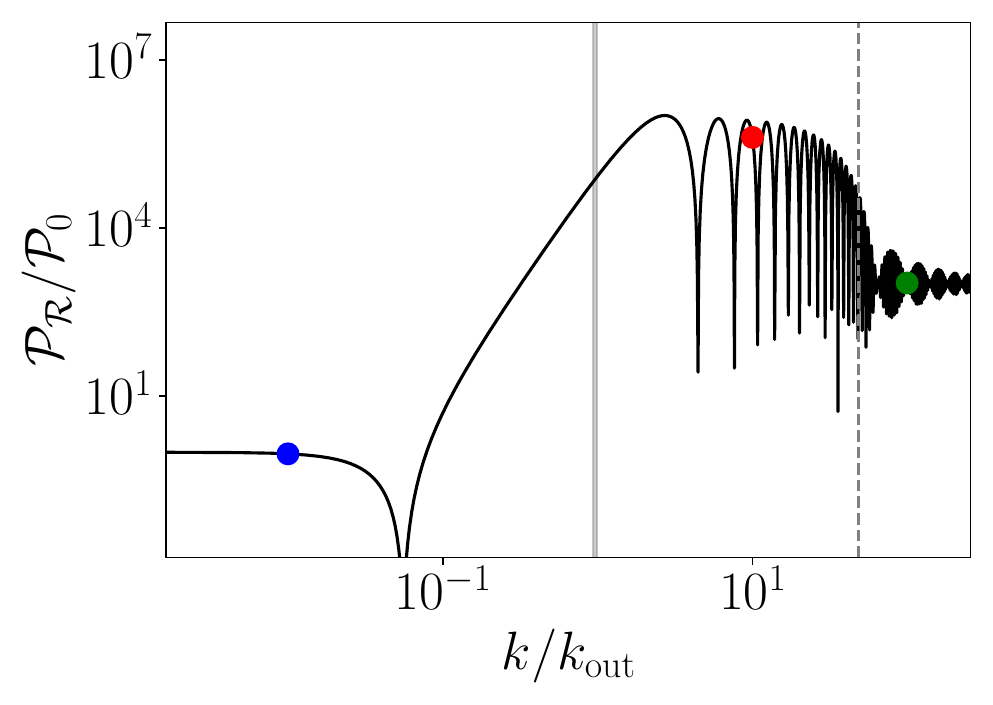}
\includegraphics[width=0.42\columnwidth]{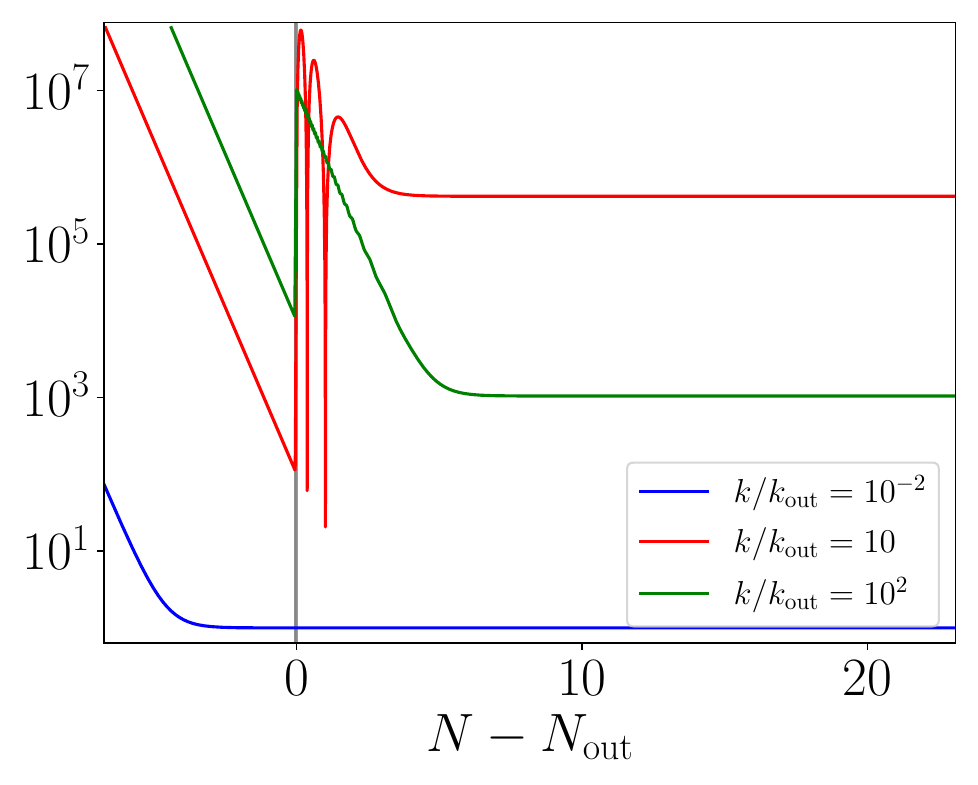}
\caption{\textit{Left panel:} linear power spectrum from a transient phase with $\eta_c = -100$, $(\epsilon_{\mathrm{in}},\epsilon_{\mathrm{out}}) = (10^{-2},10^{-5})$. The spectrum is normalized to its asymptotic value in the small-$k$ regime. The gray band is the region of the feature, while the dashed vertical line corresponds to the maximum momentum scale $k = k_{\rm out} {\cal M}/H \simeq k_{\rm out} \left\vert \eta_c \right\vert /2$ that experienced the tachyonic growth during the feature. Contrary to the physical oscillations between these two lines, coming from the excited state, the ones on the right are well understood artefacts caused by the discontinuous behaviour in \eqref{tophat}. \textit{Right panel:} time evolution (as a function of the number of \textit{e}-folds $N$, with $N = N_{\rm out}$ at the end of the feature) of the power spectrum of three representative modes highlighted on the left panel (also normalized to the final value of the power at IR scales). The modes showed with blue, red, and green color left the horizon, respectively, well before, slightly after, and well after the transition. The comparison between their powers confirms the scalings (\ref{enhance1}) and (\ref{enhance1+2}) derived in the text. 
}
\label{PSlin-etajump}
\end{figure}

Imposing $\tau_{\rm out} = \tau_{\rm in}$, or $\eta_c = 0$ in these relations results in $\alpha_k = 1$ and $\beta_k =0$, namely, these relations correctly indicate that the field remains in its vacuum state in absence of a jump in $\eta$. For a non-vanishing jump, from the Bogoliubov coefficients (\ref{alpha-jump}) and (\ref{beta-jump}) we obtain the linear power spectrum (\ref{PR-excited}) of the excited state. Let us provide a 
more immediate and physical
estimate of the main properties of the spectrum with the aid of the example shown in Figure \ref{PSlin-etajump}. In this example we choose a large and negative $\eta$, leading to a very short feature region 
\be
\tau_{\rm in} \equiv \tau_{\rm feature} - \delta \tau \;\;,\;\; 
\tau_{\rm out} \equiv \tau_{\rm feature} + \delta \tau \;\;,\;\; 
\frac{\delta \tau}{\left\vert \tau_{\rm feature} \right\vert} \ll 1 \;, 
\label{tf-dt}
\ee
in which $\epsilon$ rapidly decreases. The vertical gray band in the left panel of the figure indicates the modes that exit the horizon during this short feature phase. 

The spectrum shown in the figure reaches asymptotically constant values at the smallest and largest momenta shown. This can be mathematically understood by taking the small and large argument limits of the Hankel functions in eqs. (\ref{alpha-jump}) and (\ref{beta-jump}), that gives 

\be
\lim_{k \ll k_{\rm feature} } \mathcal{P}_\R \left( k \right) = \frac{H^2}{8 \pi^2 \Mp^2 \,\epsilon_{\rm in}} \;\;\;\;\;,\;\;\;\;\; 
\lim_{k \gg k_{\rm feature} } \mathcal{P}_\R \left( k \right) = \frac{H^2}{8 \pi^2 \Mp^2 \,\epsilon_{\rm out}} \;, 
\label{asymptotic-power-spectra}
\ee 
where $k_{\rm feature} \equiv a \left( \tau_{\rm feature} \right)  H \left( \tau_{\rm feature} \right)$ is the comoving momentum of modes that left the horizon during the feature. These two relations give the expected result that modes that leave the horizon well before and well after the feature have a standard (nearly) scale invariant linear power spectrum, inversely proportional to the slow-roll parameter $\epsilon$ evaluated at their horizon crossing times. A negative and large $\eta$ in the feature region results in $\epsilon_{\rm out} \ll \epsilon_{\rm in}$, and hence in an enhancement of the power in the asymptotic large-$k$ regime with respect to the asymptotic small-$k$ regime, 
\be
\mathcal{P}_\R \left( k \gg k_{\rm out} \right) = {\rm O } \left( \frac{\epsilon_{\rm in}}{\epsilon_{\rm out}} \right) \, \mathcal{P}_\R \left( k \ll k_{\rm in} \right) \;. 
\label{enhance1}
\ee 
This enhancement takes place for all modes $k \ga k_{\rm out}$ that exited the horizon after the feature. The tree-level power spectrum shown in Figure \ref{PSlin-etajump} indicates that modes that exited the horizon only slightly after the transition (the following discussion quantifies the meaning of ``slightly'') are actually more enhanced with respect to the IR modes than eq. (\ref{enhance1}) suggests. As we now show these modes are subject to an additional enhancement that, once combined with  (\ref{enhance1}), produces the scaling observed in the figure. To see this, let us consider  the linear equation for the Mukhanov-Sasaki variable $v_k = \sqrt{2\epsilon} a \Mp\R_k$:
\be\label{MSS}
v_k'' + \left( k ^2 - \frac{2+\mathcal{M}^2/H^2}{\tau^2}\right)v_k = 0 \;\;\;,\;\;\; 
\mathcal{M} \equiv H \sqrt{\left( \frac{\eta}{2} \right)^2 + \frac{3 \eta}{2}} \;, 
\ee
where $\eta$ has been assumed to be constant, and terms proportional to $\epsilon$ have been disregarded in $\mathcal{M}$. This is the same equation as the one of a (canonically normalized) scalar field in de Sitter spacetime with negative mass squared $-\mathcal{M}^2$. During the feature, $\mathcal{M} \simeq \frac{\vert \eta_c \vert}{2} \, H \gg H$. Namely, the adiabatic perturbation has a large tachyonic mass that renders all modes with momenta up to $\mathcal{M} / ( H \left\vert \tau_{\rm feature} \right\vert)$ highly unstable during the feature.  We can disregard the expansion of the universe in the short time $\delta \tau$ of the transient non-slow-roll evolution and solve eq. (\ref{MSS}) in this period.~\footnote{For definiteness, we set $\tau = \tau_{\rm feature}$ in eq. (\ref{MSS}), although any other time in the feature region leads to the same estimate in eq. (\ref{enhance2}) below.} This leads to the exponential enhancement of the amplitude of $v_k$, and hence of $\R_k$, 
\be
{\rm e}^{\frac{\vert \eta_c \vert}{2} \, \frac{\delta \tau}{\tau_{\rm feature}}} = 
\left( {\rm e}^{\frac{\delta \tau}{\tau_{\rm feature}}} \right)^{\frac{\vert \eta_c \vert}{2}} \simeq \left( 1 + \frac{\delta \tau}{\tau_{\rm feature}} \right)^{\frac{\vert \eta_c \vert}{2}} \simeq \left( \frac{\tau_{\rm out}}{\tau_{\rm in}} \right)^{\frac{\vert \eta_c \vert}{2}} = \sqrt{\frac{\epsilon_{\rm in}}{\epsilon_{\rm out}}} \;.  
\label{enhance2}
\ee 
Once squared, this results in a further additional power of $\frac{\epsilon_{\rm in}}{\epsilon_{\rm out}}$ with respect to eq. (\ref{enhance1}), namely to  
\be
\mathcal{P}_\R \left( k_{\rm out} < k < \vert \eta_c \vert \, k_{\rm out} / 2 \right) = {\rm O } \left( \frac{\epsilon_{\rm in}^2}{\epsilon_{\rm out}^2} \right) \, \mathcal{P}_\R \left( k \ll k_{\rm in} \right) \;. 
\label{enhance1+2}
\ee 
in agreement with Figure \ref{PSlin-etajump}. We thus confirm, and explain in these simple terms, the power spectrum enhancement found in \cite{Inomata:2021tpx} for a phase of constant and negative large $\eta$. Note moreover that the tree-level power spectrum is also enhanced at momentum scales $\lesssim$ than $\kf$, due to the resurrection of the would-be decaying mode, typical of ultra-slow-roll like setups \cite{Tsamis:2003px,Kinney:2005vj} (see also \cite{Byrnes:2018txb,Ballesteros:2020qam,Tasinato:2020vdk}).

\begin{figure}[]
\centering
\includegraphics[width=0.49\columnwidth]{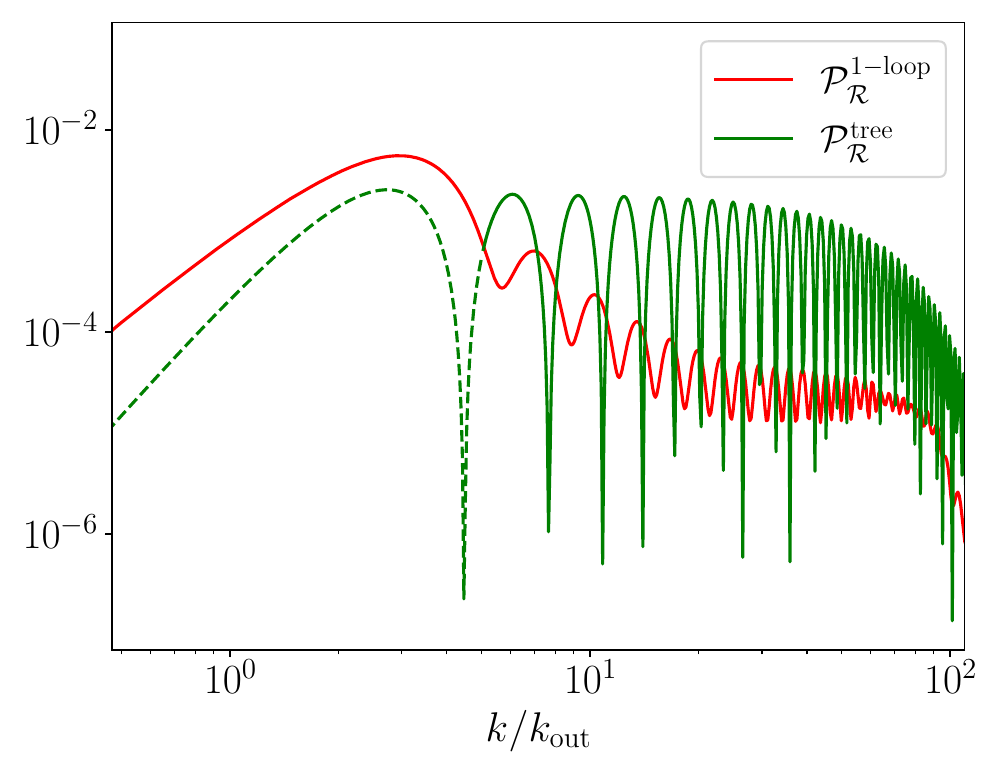}
\includegraphics[width=0.49\columnwidth]{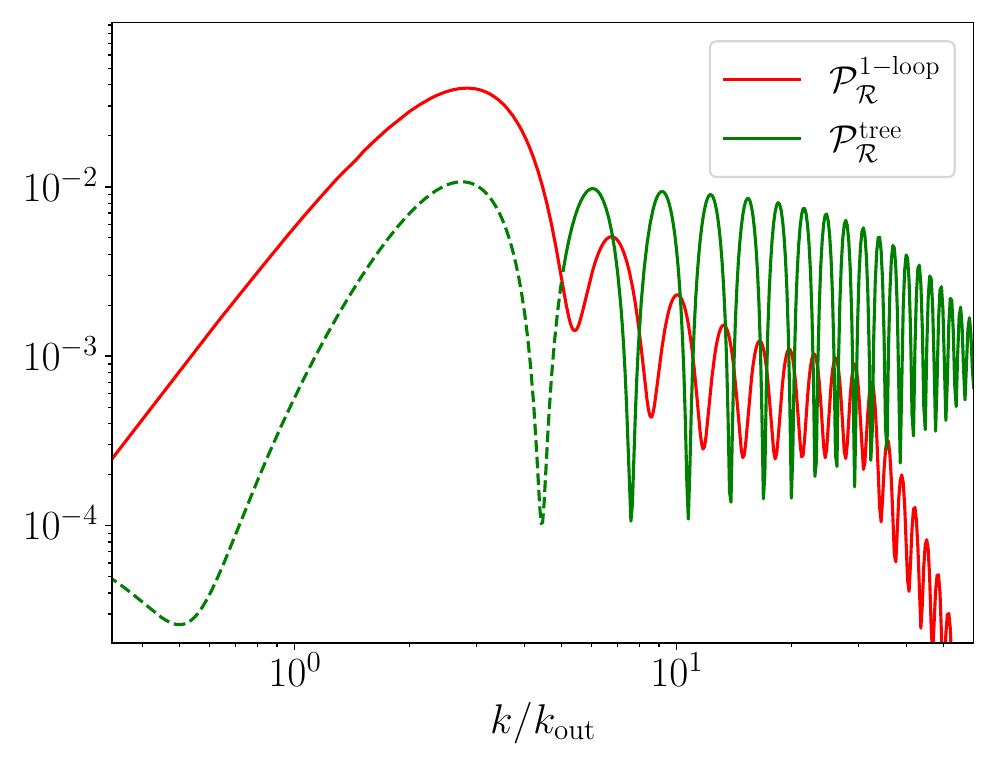}
\caption{Two examples of the power spectrum for the transient non-slow-roll single-field dynamics explained in the text, characterized by 
$\eta_c = -200, \, \epsilon_{\rm out}= 5\cdot 10^{-5}$,  $\epsilon_{\rm in} = 0.05$, $\p_0 =2.2\cdot 10^{-9}$ (left panel), and by $\eta_c = -50, \epsilon_{\rm out}= 9\cdot 10^{-3}$, $\epsilon_{\rm in} = 0.09$, $\p_0 = 10^{-4}$ (right panel). In each panel, the green and red lines show, respectively, the tree-level power spectrum and the one-loop correction to the linear power spectrum. The green line is dashed for modes for which the one-loop term dominates over the linear one. As discussed in the text, these scales 
contribute negligibly
to the one-loop power spectrum, showing a hierarchy of scales between the UV sourcing modes and the IR sourced ones.
}
\label{PS-etajump}
\end{figure}

As already remarked, the evolution of the linear modes after the feature ($\tau > \tau_{\rm out}$) is described by eq. (\ref{R123-sol}), with $\nu_{\rm out} = 3/2$ and with the Bogoliubov coefficients given by eqs. (\ref{alpha-jump}) and (\ref{beta-jump}). This evolution is the one 
considered in eq.~(\ref{eq:P-1-loop-final-self}) for the one-loop power spectrum due to the self interactions of the adiabatic perturbation $\Rh$. The resulting one-loop contribution is shown by the red line in Figure  \ref{PS-etajump} (the two panels that constitute the Figure refer to two different examples, namely to two different choices for the parameters $\epsilon_{\rm in,out}$ and $\eta_c$, as well as for the linear power spectrum $\p_0$ of the IR modes~\footnote{In the example shown in the right panel we assume a greater value of $\p_0$ than the one implied by CMB normalization. This example assumes that the feature takes place well after the CMB modes left the horizon, and that the linearized spectrum has grown to the value considered in the example at scales that left the horizon shortly before the feature.}). As compared to the previous figure, Figure  \ref{PS-etajump} considers a narrower range of modes, centered around those excited at the feature. We see that the one-loop correction exhibits a maximum at scales that left the horizon during the feature, in full analogy with what found previously in Sec. \ref{subsec:UVsource}. At these scales, the one-loop term is greater than the linear one, resulting in an enhancement of the first peak and of the IR tail preceding the peak. 

Note that, in this example, the resulting peak in the one-loop power spectrum has to compete with the enhancement of the tree-level power spectrum around the same scales $k\sim \kf$. That is a peculiarity of this single-field mechanism where the two effects cannot be disentangled. 

The fact that the one-loop correction dominates over the linear term at those scales might cast some doubt over the perturbative stability of this result. To address this, we performed the one-loop computation by including in eq. (\ref{eq:P-1-loop-final-self}) either (i) the linear modes at all scales, or (ii) artificially removing all the linear modes shown with a dashed line in the figure (as well as those with lower momenta than than those shown in the figure). Both choices result in indistinguishable results. 

The computation (ii) shows that, analogously to what emerged from the entropic case studied in the previous section, the one-loop contribution is truly a UV $\to$ IR effect, associated with the rescattering of modes that are well inside the horizon at the time of the feature. This is necessary to have the resonant growth of the sourced modes, that compensates for the fact that the one-loop term is  $\propto \epsilon_{\rm in}^2 \, \p_0^2$, while the linear power spectrum is $\propto \p_0$. The fact that the sourcing UV modes and the sourced IR modes now belong to the same field rather than to two different fields does not make any difference \textit{per se} in the computation, as the annihilation / creation operators of modes with different momenta commute, irrespective of whether the sourcing and sourced modes belong to the same or to different fields. This is the reason why the mathematical structure that controls the one-loop term is nearly identical in the two cases, cf. eq. (\ref{eq:P-1-loop-final}) for the case in which the adiabatic IR nonlinear modes are sourced by higher momenta modes of an entropic excited state, and eq.~(\ref{eq:P-1-loop-final-self}) 
for the case in which the adiabatic IR nonlinear modes are sourced by higher momenta modes of the adiabatic mode itself. The two cases are only distinguished by the presence of a different kernel function, arising from the fact that the cubic interactions are different in the two situations.

\section{Rescattering in multifield inflation}\label{sec:multifield}

\subsection{General formula}
\label{subsec:general-multi}
We now consider the situation in which, as the result of a multifield dynamics, both entropic and adiabatic fluctuations ended up in an excited state.
Let us start with the standard multifield Lagrangian for the general class of nonlinear sigma models whose target-space geometry is given by the metric $G_{IJ}$: 
\begin{equation} \label{L-background}
    \frac{\mathcal{L}}{ \sqrt{-g}}
    = -\frac{1}{2}G_{IJ}(\boldsymbol{\phi})\partial^\mu \phi ^I \partial_\mu \phi ^J - V(\boldsymbol{\phi}),
\end{equation}
where $\boldsymbol{\phi}=\left(\phi^1,\cdots,\phi^{\cal N}\right)$ and $V(\boldsymbol{\phi})$ is a generic multifield potential. Let us restrict for simplicity to two fields (${\cal N}=2$).
We conveniently make use of the adiabatic-entropic basis defined by the unit vectors $e_{\sigma}^I = \dot{\phi}^I /(G_{JK}\dot{\phi}^J\dot{\phi}^K)^{1/2}$ (tangent to the background trajectory) and $e_s^I$ (orthogonal to the trajectory) with a definite orientation being selected. 
We label field fluctuations as $Q^I$ and we fix the gauge such that the adiabatic field fluctuation is set to zero, i.e. $e_{\sigma\,I} Q^I=0$ and the spatial part of the metric reads $g_{ij}=a^2 e^{2 \R} \delta_{ij}$. The dynamics can then be described in terms of the entropic field fluctuation defined as $e_{s\,I}Q^I\equiv \F$ and $\R$, the comoving curvature fluctuation.

We consider transitions leading to an excited state for both fields $\Fh$ and $\Rh$. 
Here it is convenient to use the notation $\hat{Q}_{\lambda} = \{\hat{Q}_{\R},\hat{Q}_{\F}\}$ where $\hat{Q}_{\R}\equiv \sqrt{2\epsilon}\Mp \hat{\R}$ and $\hat{Q}_{\F}\equiv \Fh$.
Both fields are set in Bunch-Davies in the far past (before the transition) and aligned with the annihilation/creation basis $a_i$, i.e.
\be\label{BD}
\lim_{k \tau\rightarrow -\infty} \alpha_{\lambda i}(k,\tau) =\delta_{\lambda i},
\ee
with $\delta_{\lambda i}$ the Kronecker delta.
Then, along its evolution the vector  $\hat{Q}_{\lambda}$ is given in terms of $N=2$ independent solutions of the two-field system, each one labelled by the index $i$ and corresponding to the initial condition \eqref{BD}.\footnote{See for instance \cite{Nilles:2001fg,Achucarro:2010da,Fumagalli:2021mpc} for a detailed discussion about quantization of a multifield system in a FLRW background.}
In particular, interactions among adiabatic and entropic fluctuations in the quadratic Lagrangian generically break the alignment in eq. \eqref{BD} and mix, for each field operator, annihilation/creation operators associated with the \textit{in} vacuum.
After the transition the two fields are decoupled at quadratic level, i.e.~the second order action for $\R$ and $\F$ are simply given by eqs. \eqref{quadraticR} and \eqref{quadraticF}. Thus, the generic excited fluctuations can be written as ~\cite{Salopek:1988qh,GrootNibbelink:2001qt,Tsujikawa:2002qx,Weinberg:2008zzc,Achucarro:2010da} 
\be\label{field generic}
\hat{Q}_\lambda (\bp,\tau) = \frac{H}{\sqrt{2 p^ 3}}\sum_{i=1}^{2}\left[ \alpha_{\lambda i}\; \U(p\tau)+ \beta_{\lambda i} \;\U^* (p\tau)\right]a_i(\bk) +\mathrm{h.c.}(-\bk),\quad \lambda = \{\R,\F\}.
\ee
The third-order Lagrangian is the sum of the two Lagrangians \eqref{third_spectator} and \eqref{Lag-RRR}, and, therefore, the source term for the nonlinear adiabatic field is the sum of eqs. \eqref{source entropic} and \eqref{Source zeta}. Expressing these equations in terms of the field variables introduced in eq. \eqref{field generic}  leads to
\begin{align}
\hat{S}(\bk,\tau) =& \frac{1}{\Mp^2}\int  \frac{d \bp}{(2\pi)^3} \Bigg[ \sum_{\lambda = \R,\,\F} \left(f_\lambda (\bp,|\bk-\bp|)\hat{Q}_\lambda (\bp,\tau)\hat{Q}_\lambda (|\bk-\bp|,\tau)\right) \\ \nonumber 
& \quad\quad\quad\quad \quad\quad+ f_{\R'}(\bp,|\bk-\bp|)\hat{Q}'_\R(\bp,\tau)\hat{Q}'_\R(|\bk-\bp|,\tau) \Bigg] \;, 
\end{align}
where we recall that $f_{\F}$, $f_{\R'}$, and $f_{\R}$ are given, respectively, in eqs. \eqref{fF}, \eqref{fRp}, and \eqref{fR}.

Following similar steps as in the previous sections,  the one-loop rescattering formula in the multifield case can be derived and recast in a compact form as
\begin{tcolorbox}[colframe=white,arc=0pt,colback=lightgray!40]
\begin{equation}
\begin{aligned}
\hspace{-1.2em} 
\mathcal{P}^{\mathrm{1-loop}}_\R (k) &= \,  \epsilon_0^2 \p_0^2\int_0^\infty \textrm{d}x \int_{|1-x|}^{1+x} \textrm{d}y  \ \cdot \sum_{i,j}\Bigg\vert\sum_{\lambda;\,{\rm s_{1,2}=\pm}}\alpha_{\lambda\, i}^{\rm s_{1}}(xk) \alpha_{\lambda\,j}^{\rm s_{2}}(yk) \mu_\lambda(x,y)\cG(\mathrm{s_1}x,\mathrm{s_2}y,\zf)   \\
& \quad\quad\quad\quad  \quad\quad\quad\quad  \quad\quad\quad\quad 
+ \sum_{{\rm s_{1,2}=\pm}} \alpha_{\R\, i}^{\rm s_{1}}(xk) \alpha_{\R\,j}^{\rm s_{2}}(yk) \mu_{\R'}(x,y)\bar{\cG}(\mathrm{s_1}x,\mathrm{s_2}y,\zf)  \Bigg\vert^2,
\label{multi} 
\end{aligned}
\end{equation}
\end{tcolorbox}
\noindent where $\cG$ and $\bar{\cG}$ have been previously defined in \eqref{master integral} and \eqref{master integral 2}, while the coefficients $\mu_\F ,\, \mu_\R$, and $\mu_{R'}$ can be found in eqs. (\ref{muF}) and (\ref{muR-Rp}). As above, the notation $\alpha^+ \equiv \alpha ,\, \alpha^- \equiv \beta$ has been introduced to render this expression more compact. Note the non-trivial sum over field indices $\lambda$ and over indices associated with the annihilation/creation operators $i,j$. That is a consequences of the quantum mixing \eqref{field generic} which, in turn, reflects the multifield nature of the phenomenon generating the excited state. The expression \eqref{multi} is one of the main results of this paper. It encompasses the excited isocurvature and self-scattering results of Sections \ref{sec:spectator2} and \ref{sec:single} and it can be readily used to compute the one-loop power spectrum from generic excited states in terms of the corresponding \bog coefficients.

\subsection{Example: strong turn in field space}\label{subsec:examplemulti}

As an example of a dynamically generated excited state we consider the multifield enhancement mechanism first proposed in \cite{Palma:2020ejf,Fumagalli:2020adf}.
In a two-field set-up, decomposing the scalar fluctuations in components tangential and orthogonal to the background trajectory, one notices that the entropic direction may manifest a transient tachyonic instability in presence of a short phase of strongly non-geodesic motion.
Crucially, the same parameter $\etaperp$ measuring the deviation of the trajectory from a geodesic acts as a quadratic mixing between adiabatic and entropic fluctuations, i.e. $\mathcal{L}^{(2)}\sim \etaperp \dot{\R}\F$, which are thus coupled already at linear level during the $\etaperp \neq 0$ phase. In this way, the tachyonic sub-Hubble growth of the linear entropic fluctuations is concurrently transferred to $\R$. The overall effect is that, after the transition, meaning when $\etaperp\simeq 0$ and the system relaxes back to a standard quasi de Sitter evolution, the process has effectively generated an excited state for both fields $\Rh$ and $\Fh$. 
Detailed descriptions of the mechanism can be found in \cite{Palma:2020ejf,Fumagalli:2020nvq,Fumagalli:2021mpc}.
\begin{figure}[t]
\begin{center}
\includegraphics[width=0.49\textwidth]
{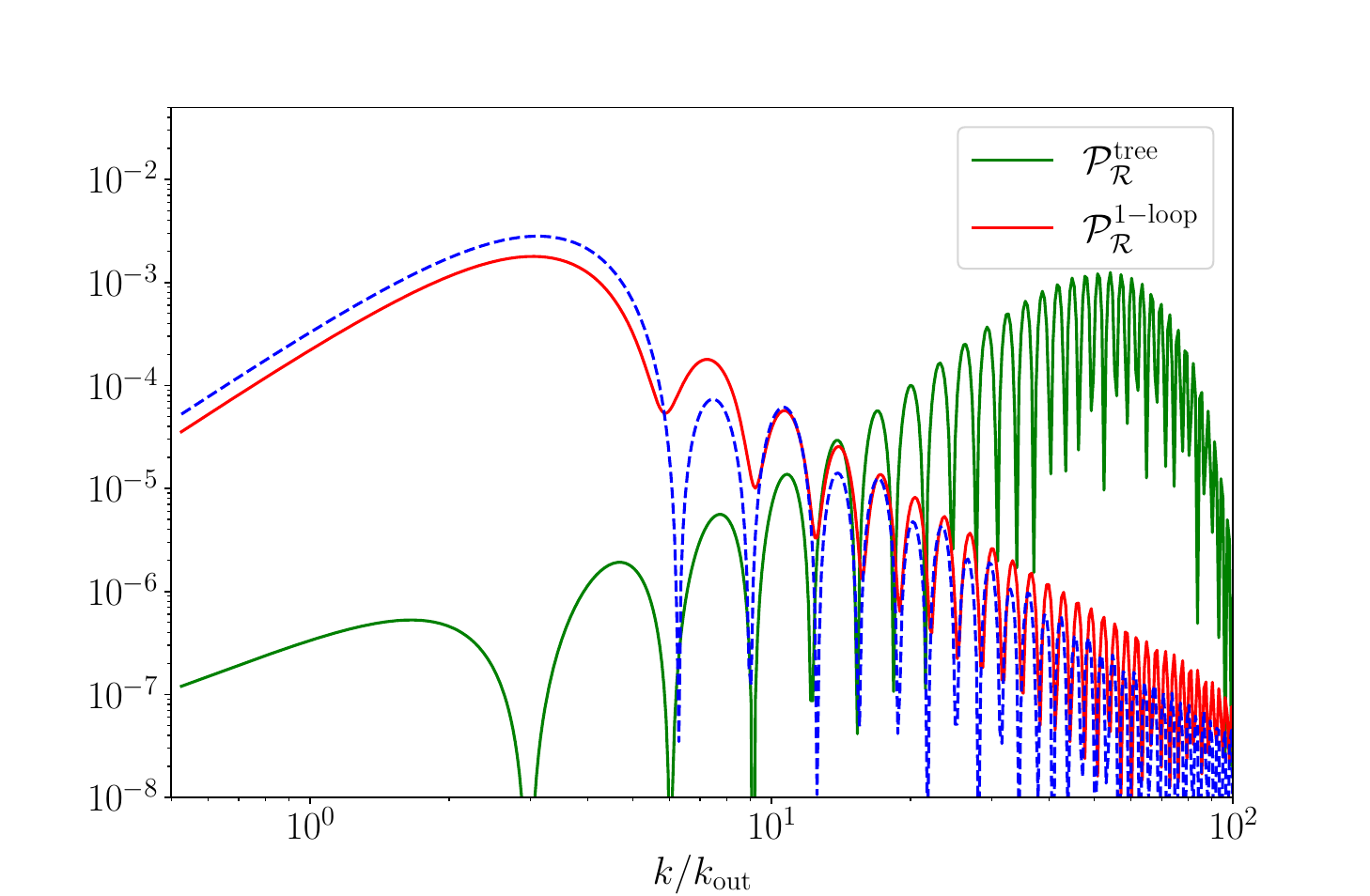}
\includegraphics[width=0.485\textwidth]
{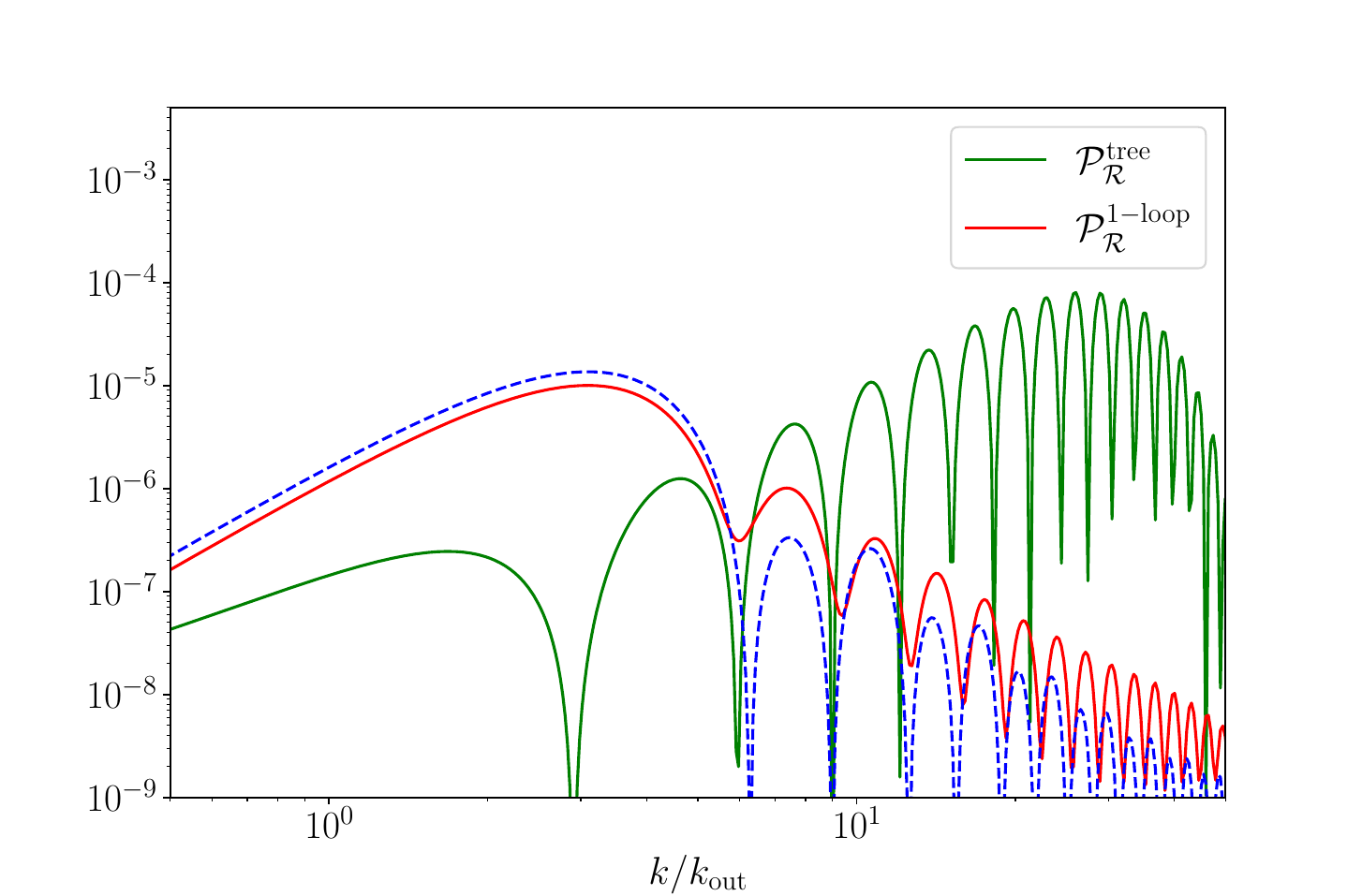}
\end{center}
\caption{Tree-level power spectrum (green curve) and one-loop power spectrum (red curve) (using eq.~\eqref{multi}) for a model with a strong turn in the  field-space trajectory, with parameters $(\etaperp,\delta)=(56,0.125)$, $\p_0 = 2.2\times 10^{-9}$ and $\epsilon _0 = 0.01$ (left panel) and $(\etaperp,\delta)=(30,0.2)$, $\p_0 = 10^{-9}$, $\epsilon _0 = 0.06$ (right panel). Blue dashed curve is the analytical approximation of eq.~\eqref{multi} derived assuming peaked Bogoliubov coefficients given in eq.~\eqref{Gauss}, following similar steps as in Sec.~\ref{subsec:analytic-peaked-specfield}. }
\label{multi_benchmark}
\end{figure}

In the simple parametrization used in these works in order to  obtain an analytical understanding, the bending parameter $\etaperp (N)$ has a top-hat time-dependence, with height and width in \textit{e}-folds given by the constants $\etaperp$ and $\delta$.
The scales which are most affected by the turn are the ones of energy $H_{\rm feature}\etaperp$, i.e., the modes around $k_*=k_{\rm feature}\etaperp$. Thus, $\gamma = k_*/k_{\rm out}=\etaperp e^{-\delta /2}$, which, for sharp turns $\delta \ll 1$ is $\gamma \simeq \etaperp$.
Considering the situation in which the entropic field $\F$ is massless during the turn ($\xi=-3$ in the notations of \cite{Fumagalli:2020nvq,Fumagalli:2021mpc}), the \bog coefficients read 
\begin{align}
\alpha _{\R 1}&= -\frac{e^{\etaperp \delta \, S}}{4S}, \, \, \, \, \, \, \, \, \alpha _{\R 2}=\frac{ie^{\etaperp \delta\,  S}}{4S},\nonumber \\
\alpha _{\F i}&= i\frac{S^2+\kappa ^2}{2\kappa}\alpha _{\R i}, \nonumber \\
\beta _{\R i}&= -e^{i\theta _k}\alpha _{\R i}, \, \, \, \, \, \, \, \, \beta _{\F i}=e^{i\theta _k}\alpha _{\F i},
\label{bog-turn}
\end{align}
where
\be
S=\sqrt{2\kappa -\kappa ^2}, ~{\rm and} \, \, \, \, \, \, \, \, \theta _k = 2e^{-\delta /2}\kappa \etaperp + 2 \arctan (\kappa /S).
\ee
Using these in eq. \eqref{multi}, one obtains the results in Figure \ref{multi_benchmark}. As discussed in Subsection \ref{subsec:UVsource}, the requirement that the excitation of the fields is described by a weakly coupled theory bounds the parameter of specific models. Tentative bounds for strong sharp turns considered here have been discussed in \cite{Fumagalli:2020nvq,Fumagalli:2021mpc} and conservatively read $\etaperp^4 \, {\cal P}_{\R}^{\textrm{max}} \lesssim 1-100$. For illustrative purposes, in Figure \ref{multi_benchmark}, we consider two distinct possibilities, one of which violates this tentative bound (on the left), while the other one is marginally compatible with it (on the right). In this framework, and contrary to what happened in the single-field context, UV linear modes are the only ones enhanced by the mechanism (in contrast to modes of the same size of the horizon at that time), since $\epsilon$ and $\eta$ are kept trivial throughout the transition. As a net result, in this context, the IR one-loop peak, which is by now a familiar result of the rescattering process highlighted in this work, can prominently emerge as a second distinct peak, for both examples considered in the figure.

\section{Conclusions}\label{sec:conclusion}
A growing experimental effort is underway to explore the late stages of inflation at smaller scales than the CMB and LSS. Any significant deviation from scale invariance that may be uncovered by these experiments would indicate the breaking of the approximate time translation symmetry that characterizes inflationary physics at the CMB scales~\cite{Planck:2018jri}, and point to the emergence of new dynamics. The latter could involve interactions between the inflaton and some other fields that become suddenly relevant at some specific stage during inflation, sharp turns in multi-field space, sharp features in the potential with  localized deviation from the standard slow-roll evolution, or other possibilities that have been extensively studied in recent literature. These effects can give rise to sharp peaks in the power spectrum of primordial perturbations, potentially leading to observable consequences in CMB distortions, gravitational-wave observatories, and primordial black hole searches.

Most studies in this field focus on computing the growth of power at the linearized level in perturbation theory. In this work, we instead address the question of whether the enhancement of primordial perturbations at a particular scale $k_*$ also leads to significant nonlinear effects. Since there are various mechanisms that can generate this enhancement, the quantitative answer to this question depends on the specific inflationary model employed. In order to draw some general conclusions, we only consider the universal gravitational interactions between the different modes after the enhancement has been established, while disregarding any additional model-specific couplings. These working assumptions are likely to provide a conservative estimate of the nonlinear effects.

Furthermore, we assume that the scale $k_*$ experiences a significant enhancement, reaching a regime of large occupation number while still inside the horizon. We demonstrate that immediately after their creation, these enhanced modes undergo rescattering (nonlinear interactions) with each other. As a result, they act as a source for the primordial curvature perturbations at scales below $k_*$, down to the horizon scale at that time $k_{\mathrm{IR},\mathrm{near}}$. We quantify this infrared enhancement by attributing it to the constructive resonance between positive and negative frequency modes. The magnitude of this effect, which we evaluate using the dominant one-loop diagram, can be analytically understood and it is related to the ratio between $k_*$ and the comoving Hubble rate at the time of the enhancement. 

We consider three distinct cases: (i) when the excited state is an isocurvature field, (ii) when it is the curvature perturbation itself, and (iii) when it is a combination of curvature and isocurvature fluctuations. For each case, we derive an analytic expression for the one-loop correction in terms of the Bogoliubov coefficients that characterize the   excited state. We also present a simplified general formula that yields a concise and immediate expression for the one-loop correction when the spectrum of enhanced modes is characterized by a narrow peak.
We evaluate this analytic expression using specific examples from models considered in the existing literature. Our results demonstrate that the IR one-loop contribution to the curvature power spectrum can ``emerge" alongside the ``primary" tree-level ultraviolet (UV) peak. This can be observed in Figures \ref{PS-etajump} and \ref{multi_benchmark}. 

Due to the dominance of the one-loop contribution over the tree-level contribution at these near-infrared (IR) scales, it is legitimate to wonder whether our results are under perturbative control. Fortunately, the separation between the sourcing ($k_*$) and the sourced ($k_{\mathrm{IR},\mathrm{near}}$) scales allows us to examine this issue more closely. To illustrate this point, we can refer to the two panels in Figure \ref{PS-etajump}, where the tree-level and one-loop contributions dominate at different momentum scales. We computed the one-loop contribution using two different approaches: (i) by consistently including all linear modes, and (ii) by artificially excluding from the loop momentum integral (which is performed over the sourcing modes) the IR modes where the tree-level contribution was found to be subdominant in the previous calculation. Remarkably, the two computations yielded indistinguishable results for the one-loop contribution. In other words, we verified that only the scales for which the one-loop contribution is well below the tree-level contribution are responsible for the rise in power observed in the one-loop computation at the $k_{\mathrm{IR},\mathrm{near}}$ scales.

This test serves to ensure that the infrared (IR) rescattering phenomenon is not merely an artifact resulting from an inadequate modeling of the sourcing field. This verification is sufficient for our objectives, as previously mentioned, which aim to demonstrate that a substantial excitation at sub-horizon scales generally leads to a significant transfer of power to $k_{\mathrm{IR},\mathrm{near}}$ scales. It might be possible that higher-loop contributions may also need to be taken into account to fully quantify the magnitude of this infrared effect. However, there is no reason to expect that these higher-loop terms will cancel out the sourcing effects of the UV modes $k_*$ (where the one-loop term already provides a negligible contribution). More generally,  while our working assumptions enable us to obtain a conservative estimate of the effect, precise quantitative results will require a thorough examination of the full set of interactions within each specific model, on a case-by-case basis. 

Curvature perturbations of sufficiently high amplitude act as seeds of PBH at horizon re-entry. As we saw, a large excitation of the scalar perturbations will be necessarily accompanied by a significant transfer of power in the IR. The full amplitude of scalar modes is the quantity to be used for the PBH abundance computations, so that our results, when significant, can have a sizeable impact on the resulting PBH amount and mass distribution. Scalar modes also source gravitational waves at horizon re-entry. Re-entries occurring during the radiation dominated era result in a GW signal with fractional energy density per unit $\log k$ of amplitude $h^2 \Omegagw^{\textrm{rad}} \simeq 10^{-5} {\cal P}_{\R}^2$ \cite{Acquaviva:2002ud, Mollerach:2003nq, Ananda:2006af, Baumann:2007zm}.\footnote{This result disregards the connected contribution to the scalar four-point function sourcing the GW spectrum, see \cite{Garcia-Bellido:2017aan,Unal:2018yaa,Atal:2021jyo,Adshead:2021hnm,Garcia-Saenz:2022tzu,Garcia-Saenz:2023zue} for works discussing the latter effect.} This signal adds up to the one generated already during inflation, $h^2 \Omegagw^{\textrm{inf}} \simeq 6 \cdot 10^{-7} {\cal P}_t$. In this paper we have computed the resonant growth of the scalar curvature perturbation considering only gravitational interactions. The excited linear modes also resonantly source GW~\cite{Fumagalli:2021mpc}. We verified that, not surprisingly, due to the universality of gravity, these two resonant productions are comparabable, namely $\mathcal{P}^{\mathrm{1-loop}}_\R \sim {\cal P}^{\mathrm{1-loop}}_t$. From this, the amount of GW sourced at re-entry by the IR scalar modes vs. those resonantly produced during inflation scales as $ \Omegagw^{\textrm{rad}}/\Omegagw^{\textrm{inf}} \sim 10^2 \mathcal{P}^{\mathrm{1-loop}}_\R$. The GW signature from the IR rescattering that we have computed can therefore emerge only for a very large scalar power spectrum, as it might be the case for scenarios leading to significant PBH abundance. As we mentioned above, it is well conceivable that the additional direct interactions in the scalar sector present in concrete models can produce an IR scalar rescattering (and an associated GW production at re-entry) greater than the one obtained here. More in general, we believe that our general lower limit estimates of the IR rescattering provide a compelling motivation for further research in this topic.

\section*{Acknowledgments}
The authors thank Keisuke Inomata for useful discussions. The research of J.F. is supported by the State Agency for Research of the Spanish Ministry
of Science and Innovation through the “Unit of Excellence María de Maeztu 2020-2023” award to the Institute of Cosmos Sciences (CEX2019-000918-M) and by grants 2021-SGR00872 and PID2019-105614GB-C22.
S.B. is supported by the ``Progetto di Eccellenza'' of the Department of Physics and
Astronomy of the University of Padua. S.B. and M.P. acknowledge support by Istituto Nazionale
di Fisica Nucleare (INFN) through the Theoretical Astroparticle Physics (TAsP) project. S.RP is supported and L.T.W has been supported by the European Research Council under the European Union's Horizon 2020 research and innovation programme (grant agreement No 758792, Starting Grant project GEODESI), as well as the IEA Gravitational-Wave Primordial Cosmology funded by CNRS. This article is distributed under the Creative Commons Attribution International Licence (\href{https://creativecommons.org/licenses/by/4.0/}{CC-BY 4.0}).

\bibliographystyle{JHEP}
\bibliography{Biblio}

\end{document}